\documentclass[preprint,12pt]{elsarticle}
\usepackage{graphics}
\usepackage{epsfig}
\usepackage{amssymb}
\usepackage{multirow}
\usepackage{color}

\usepackage{nfontdef}

\newcommand{\ignore}[1]{}

\newcommand{\vek}[1]{\mathchoice{\displaystyle\boldsymbol#1}
{\textstyle\boldsymbol#1}{\scriptstyle\boldsymbol#1}
{\scriptscriptstyle\boldsymbol#1}}

\usepackage{array}
\newcolumntype{L}[1]{>{\raggedright\let\newline\\\arraybackslash\hspace{0pt}}m{#1}}
\newcolumntype{C}[1]{>{\centering\let\newline\\\arraybackslash\hspace{0pt}}m{#1}}
\newcolumntype{R}[1]{>{\raggedleft\let\newline\\\arraybackslash\hspace{0pt}}m{#1}}

\journal{Applied Mathematics and Computation}

\begin{document}

\newpage
%%%%%%%%%%%%%%%%%%%%%%%%%%%%%%%%%%%%%%%%%%%%%%%%%%%%%%%%%%%%%%%%
\begin{frontmatter}

% Title, authors and addresses
\title{Uncertainty Propagation in Elasto-Plastic Material}

\author[ctu]{Jan S\'{y}kora}
\ead{jan.sykora.1@fsv.cvut.cz}

\author[ctu]{Anna Ku\v{c}erov\'a\corref{auth}}
\ead{anicka@cml.fsv.cvut.cz}

\cortext[auth]{Corresponding author. Tel.:~+420-2-2435-5326;
fax~+420-2-2431-0775}

\address[ctu]{Department of Mechanics, Faculty of Civil Engineering,
  Czech Technical University in Prague, Th\'{a}kurova 7, 166 29 Prague
  6, Czech Republic}

%%%%%%%%%%%%%%
\begin{abstract}
Macroscopically heterogeneous materials, characterised mostly by
comparable heterogeneity lengthscale and structural sizes, can no
longer be modelled by deterministic approach instead. It is
convenient to introduce stochastic approach with uncertain
material parameters quantified as random fields and/or random
variables. The present contribution is devoted to propagation of
these uncertainties in mechanical modelling of inelastic
behaviour. In such case the Monte Carlo method is the traditional
approach for solving the proposed problem. Nevertheless,
convergence rate is relatively slow, thus new methods (e.g.
stochastic Galerkin method, stochastic collocation approach, etc.)
have been recently developed to offer fast convergence for
sufficiently smooth solution in the probability space. Our goal is
to accelerate the uncertainty propagation using a polynomial chaos
expansion based on stochastic collocation method. The whole
concept is demonstrated on a simple numerical example of uniaxial
test at a material point where interesting phenomena can be
clearly understood.
\end{abstract}


\end{frontmatter}

\section{Introduction}
Probabilistic or stochastic mechanics deals with mechanical
systems, which are either subject to random external influences -
a random or uncertain environment, or are themselves uncertain, or
both, cf. e.g. the
reports~\cite{Gutierrez:2004,Keese:2003,Matthies:2007}. From a
mathematical point of view, these systems can be characterised by
stochastic ordinary/partial differential equations (SODEs/SPDEs),
which can be solved by stochastic finite element method (SFEM).
SFEM is an extension of the classical deterministic finite element
approach to the stochastic framework i.e. to the solution of
stochastic (static and dynamic) problems involving finite elements
whose properties are random, see~\cite{Stefanou:2009}

Nowadays, Monte Carlo (MC) is the most widely used technique in
simulating models driven by SODEs/SPDEs. MC simulations require
thousands or millions samples because of relatively slow
convergence rate, thus the total cost of these numerical
evaluations quickly becomes prohibitive. To meet this concern, the
surrogate models based on the polynomial chaos expansion (PCE),
see~\cite{Wiener:1938,Xiu:2002}, were developed as a promising
alternative. PC-based surrogates are constructed by different
fully-, semi- or non-intrusive methods based on the stochastic
Galerkin method~\cite{Ghanem:2012,Matthies:2007}, stochastic
collocation (SC) method~\cite{Babuska:2004,Babuska:2007,Xiu:2009}
or DoE (design of experiments)-based linear regression
~\cite{Blatman:2010}. The principal differences among these
methods are as follows. Stochastic Galerkin method is purely
deterministic (nonsampling method), but leads to solution of large
system of equations and needs a complete intrusive modification of
the numerical model (and/or existing finite element code) itself.
Consequently, suitable robust numerical solver is required. On the
other hand, SC method is a sampling method, does not require
intrusive modification of a model, but uses a set of model
simulations. The computation of PCE coefficients is based on
explicit formula and computational effort depends only on the
chosen level of accuracy and corresponding number of grid points,
see~\cite{Ghanem:2012,Matthies:2007,Xiu:2005}. The linear
regression is based again on a set of model simulations performed
for a stochastic design of experiments, usually obtained by Latin
Hypercube Sampling. The PCE coefficients are then obtained by a
regression of model results at the design points, which leads to a
solution of a system of equations. A short overview of yet another
approaches for solving SODEs/SPDEs is available
in~\cite{Xiu:2005}.

In particular, this paper is focused on the modelling of
uncertainties in elasto-plastic material. While numerical studies
using MC methods have been presented during several years, the
PC-based strategies has emerged only recently,
~\cite{Anders:1999,Arnst:2012,Rosic:2008}. The authors mostly
extended original work of Ghanem and Spanos~\cite{Ghanem:2012} to
elasto-plasticity problem by approximating spatial varying
material properties and model responses using Karhunen-Lo\`{e}ve
and PCE, respectively. The interested reader may also consult an
excellent work on this subject by Rosic~\cite{Rosic:2012}.
However, all these works concentrate especially on the stochastic
Galerkin method and SC method is discussed only marginally. Hence
this paper is devoted to the application of SC method and for a
sake of simplicity we focus here on uncertainty propagation in
elasto-plastic material at a single material point.

The paper is organised as follows: A problem setting is presented
in Section~\ref{sec:ps}, followed by description of material model
in Section~\ref{sec:mm} and surrogate model in
Section~\ref{sec:sm}. Section~\ref{sec:exam} then demonstrates the
proposed framework on elasto-plasticity problem at material point.
The essential findings are summarised in Section~\ref{sec:con}.

\section{Problem setting}
\label{sec:ps}

This paper is focused on the modelling of uncertainties in
properties of elasto-plastic material and investigates the
influence of such uncertainties on mechanical behaviour. To
fulfill this objective, we introduce a bounded body
$D\subset\D{R}^{3}$ (reference configuration) with a piecewise
smooth boundary $\partial D$. In particular, the Dirichlet and
Neumann boundary conditions are imposed on
$\Gamma_{\mathrm{D}}\subset\partial D$ and
$\Gamma_{\mathrm{N}}\subset\partial D$, respectively, such that
$\partial D=\Gamma_{\mathrm{D}}\cup\Gamma_{\mathrm{N}}$. Moreover,
we are interested in the time-dependent behavior of $D$, thus we
consider a time interval $[0,T]\subset\D{R}_{+}$. The evolution of
the material body $D$ in the geometrically linear regime is
expressed as
\begin{equation}
\vek{u}:D\times[0,T]\longrightarrow \D{R}^{3}, \label{eq:ps1}
\end{equation}
where $\vek{u}\,\mathrm{[m]}$ is the displacement field. In a
quasi-static setting, the linear momentum balance equation is then
described by
\begin{eqnarray}
-\,\mathrm{div}\vek{\sigma}(x,t)=\vek{f}(x,t), \quad x\in D,
\,t\in [0,T]
\end{eqnarray}
and corresponding boundary conditions
\begin{eqnarray}
\vek{\sigma}(x,t)\cdot\vek{n}(x)& = & \vek{t}_{\mathrm{N}}(x,t),
\qquad\qquad\qquad\qquad \nonumber\\
&& x\in
\Gamma_{\mathrm{N}}, \,t\in [0,T], \\
\vek{u}(x,t)&=&\vek{u}_{\mathrm{D}}(x,t), \qquad\qquad\qquad\qquad
\nonumber\\
&& x\in \Gamma_{\mathrm{D}}, \,t\in [0,T],
\end{eqnarray}
where $\vek{\sigma}(x,t)\,\mathrm{[Pa]}$ is the stress tensor,
$\vek{f}(x,t)\,\mathrm{[Nm^{-3}]}$ is the body forces,
$\vek{n}\,\mathrm{[-]}$ is the exterior unit normal,
$\vek{t}_{\mathrm{N}}(x,t)\,\mathrm{[Pa]}$ is the prescribed
surface tension and $\vek{u}_{\mathrm{D}}(x,t)\,\mathrm{[m]}$ is
the prescribed displacement.

Consider now a system involving material variability. If the input
parameter is defined as a random variable and/or field, the system
would be governed by a set of SPDEs and the corresponding
responses would also be random vectors of nodal displacements, see
~\cite{Matthies:2005,Kucerova:2013}. Let $(\Omega
,\mathscr{S},\D{P})$ be a complete probability space with $\Omega$
the set of the elementary events $\omega$, $\D{P}$ the probability
measure and $\mathscr{S}$ an $\sigma$-algebra on the set $\Omega$.
Following previous definitions of the evolution of the material
body $D$ (Eq.~\ref{eq:ps1}), we are now concerned with the mapping
in the stochastic setting:
\begin{equation}
\vek{u}:D\times[0,T]\times\Omega\longrightarrow \D{R}^{3}.
\end{equation}
Consequently, the linear momentum balance equation is then given
by
\begin{eqnarray}
&& -\,\mathrm{div}\vek{\sigma}(x,t,\omega)=\vek{f}(x,t,\omega), \qquad\qquad\qquad\qquad \nonumber\\
&& \qquad\qquad\qquad x\in D, \,t\in [0,T], \,\omega\in \Omega
\end{eqnarray}
and corresponding boundary conditions
\begin{eqnarray}
\vek{\sigma}(x,t,\omega)\cdot\vek{n}(x)& = &
\vek{t}_{\mathrm{N}}(x,t,\omega),
\qquad\qquad\qquad\qquad \nonumber \\
&& x\in \Gamma_{\mathrm{N}}, \,t\in [0,T], \,\omega\in \Omega,\\
\vek{u}(x,t,\omega)&=&\vek{u}_{\mathrm{D}}(x,t,\omega),
\qquad\qquad\qquad\qquad \nonumber
\\
&& x\in \Gamma_{\mathrm{D}}, \,t\in [0,T], \,\omega\in \Omega.
\end{eqnarray}
In order to solve this stochastic partial differential equation
and obtain the approximate responses of the system, MC method is
usually used. The effort of performing MC simulations is high, and
hence strategies based on the surrogate models have been developed
to accelerate the SPDEs solution. In particular, we employ the
polynomial expansion and collocation method described concisely in
Section ~\ref{sec:sm}.

\section{Material Model}
\label{sec:mm}

In order to demonstrate a performance of the SC method, we briefly
introduce the mathematical formulation of the deterministic
elasto-plastic behaviour. The basic equations of flow plasticity
theory start from the decomposition of strain rate vector
$\dot{\vek{\varepsilon}}\,\mathrm{[-]}$ in an elastic (reversible)
part $\dot{\vek{\varepsilon}}\,\mathrm{[-]}$ and a plastic
(irreversible) part
$\dot{\vek{\varepsilon}}_{\mathrm{p}}\,\mathrm{[-]}$, see
~\cite{Grassl:2006,Lourenco:1997},
\begin{equation}
\dot{\vek{\varepsilon}} = \dot{\vek{\varepsilon}}_{\mathrm{e}} +
\dot{\vek{\varepsilon}}_{\mathrm{p}}.
\end{equation}
The elastic strain rate is related to the stress rate according to
constitutive relation for an isotropic elastic material as
\begin{equation}
\dot{\vek{\sigma}} = \vek{D}_{\mathrm{e}} :
\dot{\vek{\varepsilon}_{\mathrm{e}}},
\end{equation}
where $\vek{D}_{\mathrm{e}}\,\mathrm{[Pa]}$ is the elastic
material stiffness matrix. The associated flow rule is usually
defined using the plastic multiplier and the plastic potential,
which is in this case equal to a particular yield criterion, as
\begin{equation}
\dot{\vek{\varepsilon}}_{\mathrm{p}} = \dot{\lambda}\frac{\partial
f(\vek{\sigma},\sigma_{\mathrm{y}})}{\partial \vek{\sigma}}.
\end{equation}
It remains to characterise loading/unloading conditions
established in standard Karush--Kuhn--Tucker form as
\begin{equation}
f\leq 0, \quad \dot{\lambda} \geq 0, \quad \dot{\lambda}f = 0.
\end{equation}
Just for the sake of completeness, we introduce the simplest and
most useful yield condition formulated by {\it
Maxwell--Huber--Hencky--von Mises}, often called $J_2$-plasticity,
see~\cite{Dunne:2005}. Here, the yield criterion is expressed as
\begin{equation}
f(\vek{\sigma},\sigma_{\mathrm{y}}) = \sqrt{J_2} -
\frac{\sigma_{\mathrm{y}}(\kappa)}{\sqrt{3}},
\end{equation}
where $J_2\,\mathrm{[Pa^2]}$ is the second invariant of the
deviatoric stress, $\sigma_{\mathrm{y}}(\kappa)\,\mathrm{[Pa]}$ is
the tensile yield strength. Here, we assume a bilinear form of
strain hardening plasticity described by an evolution of the
tensile yield strength as a function of a hardening parameter
$\kappa$ as
\begin{equation}
\sigma_{\mathrm{y}}(\kappa) =
\sigma_{\mathrm{y}}(\varepsilon_{\mathrm{p}}^{\mathrm{eq}}) =
\sigma_{\mathrm{y,0}} + H\varepsilon_{\mathrm{p}}^{\mathrm{eq}},
\end{equation}
where $H\,\mathrm{[Pa]}$ is the hardening parameter,
$\sigma_{\mathrm{y,0}}\,\mathrm{[Pa]}$ is the initial yield
strength and
$\varepsilon_{\mathrm{p}}^{\mathrm{eq}}\,\mathrm{[-]}$ is the
equivalent plastic strain calculated as
$\varepsilon_{\mathrm{p}}^{\mathrm{eq}} = \sqrt{(\frac{2}{3} \,
\vek{\varepsilon}_{\mathrm{p}}:\vek{\varepsilon}_{\mathrm{p}})}$.
The most popular approach for integrating the constitutive
equations of isotropic hardening $J_2$-plasticity is the radial
return method proposed by Krieg and Krieg, see~\cite{Krieg:1977}.
The stability and efficiency of the algorithms and also details of
the algorithmic formulation may be found
in~\cite{Dunne:2005,Horak:2009}.

\section{Surrogate model}
\label{sec:sm}

The construction of a surrogate of the computational model can be
used for a significant acceleration of each sample evaluation.
Here we use the SC method
\cite{Babuska:2004,Babuska:2007,Xiu:2009} to construct the
surrogate model based on the PCE.

\subsection{Polynomial chaos expansion}

In modelling of heterogeneous material, some material parameters
are not constants, but can be described as random variables (RVs),
namely real-valued random variables $X:\Omega\rightarrow \D{R}$
specified completely by their cumulative distribution functions
(CDFs). From a mathematical and computational point of view, it is
better to use independent random variables for numerical
integration over the probability space $\Omega$,
see~\cite{Matthies:2005,Matthies:2007}. Therefore, we introduce
set of independent Gaussian random variables $\vek{\xi}(\omega) =
(\xi_1(\omega), \dots, \xi_{s}(\omega)])^{\mrm{T}}$ with zero mean
and unit variance, see~\cite{Matthies:2007,Xiu:2002}~\footnote{Due
to positive values of some material properties, it is convenient
to also define lognormal transformation of Gaussian RV as $
q(\omega) = \exp ( \mu_g + \sigma_g \xi(\omega))$. The statistical
moments $\mu_g$ and $\sigma_g$ can be transformed from statistical
moments $\mu_q$ and $\sigma_q$ given for lognormally distributed
material property, see~\cite{Kucerova:2012}.}. According to the
Doob-Dynkin lemma~\cite{Xiu:2005}, the model response
$\vek{u}(\vek{\xi}(\omega))= \left(
  \dots, u_l(\vek{\xi}(\omega)), \dots \right)^{\mrm{T}}$ is a random
vector which can be expressed in terms of the same random
variables $\vek{\xi}(\omega)$. Since $\vek{\xi}(\omega)$ are
independent standard Gaussian RVs, Wiener's PCE based on
multivariate Hermite polynomials~\footnote{We assume the full PC
expansion, where number of polynomials $r$ is fully determined by
the degree of polynomials $p$ and number of random variables $s$
according to the well-known relation $r =
(s+p)!/(s!p!)$.}---orthogonal in the Gaussian
measure---$\{H_\alpha(\vek{\xi}(\omega))\}_{\alpha \in \C{J}}$ is
the most suitable choice for the approximation
$\tilde{\vek{u}}(\vek{\xi}(\omega))$ of the model response
$\vek{u}(\vek{\xi}(\omega))$ \cite{Xiu:2002}, and it can be
written as
\begin{equation}
\tilde{\vek{u}}(\vek{\xi}(\omega)) = \sum_{\alpha \in \C{J}}
\vek{u}_\alpha
 H_{\alpha}(\vek{\xi}(\omega)),
\label{eq:Tapprox}
\end{equation}
where $\vek{u}_\alpha$ is a vector of PC coefficients and the
index set $\C{J} \subset \D{N}_0^{(\D{N})}$ is a finite set of
non-negative integer sequences with only finitely many non-zero
terms, i.e.\ multi-indices, with cardinality $|\C{J}| = R$.

\subsection{Stochastic collocation}

As a preamble, it has been proven that for sufficiently smooth
solution in the probability space, SC method achieves as fast
convergence as stochastic Galerkin method, see~\cite{Xiu:2005}.
Moreover, utilisation of existing deterministic solvers for
repetitive runs and especially no need for numerical model
modification are important practical aspects. Such properties make
the SC method more preferred alternative to stochastic Galerkin
method and MC method for solving SODEs/SPDEs,
see~\cite{Xiu:2002,Xiu:2005,Xiu:2009}.

In principle, SC method is based on the approximation of
stochastic solution using appropriate multivariate polynomials.
According to ~\cite{Xiu:2005}, the SC method can be seen as a
high-order "deterministic sampling method." The formulation is
based on an explicit expression of the PC coefficients as,
see~\cite{Babuska:2004,Babuska:2007,Xiu:2009}:
\begin{equation}
  %u_{\alpha,l} = \int_{\Omega} u_l(\vek{\xi}) H_\alpha(\vek{\xi}) \, \mathrm{d} \D{P} (\vek{\xi}) \, ,
  u_{\alpha,l} = \int_{\Omega} u_l(\vek{\xi}) H_\alpha(\vek{\xi}) \, \mathrm{d} \D{P} (\vek{\xi}) \, ,
\label{eq:sc}
\end{equation}
which can be solved numerically using an appropriate integration
(quadrature) rule on $\D{R}^{s}$. Equation \eqref{eq:sc} then
becomes
\begin{equation}
 % u_{\alpha,l} = \frac{1}{\gamma_{_\alpha}}\sum_{j=1}^{i} u_l(\vek{\xi}_j) H_\alpha(\vek{\xi}_j) w_j \, ,
  u_{\alpha,l} = \frac{1}{\gamma_{_\alpha}}\sum_{j=1}^{i} u_l(\vek{\xi}_j) H_\alpha(\vek{\xi}_j) w_j \, ,
\label{eq:sc1}
\end{equation}
where $\gamma_{_\alpha} = \D{E}[H_\alpha^2(\vek{\xi})]$ are the
normalization constants of PC basis, $\vek{\xi}_j$ stands for an
integration node, $w_j$ is a corresponding weight and $i$ is the
number of quadrature points.  Once we have obtained accurate PC
approximation, following analytical relations replacing exhaustive
sampling procedure of PC expansion are used to calculate mean
$\mu_l$ and standard deviation $\sigma^{\mathrm{STD}}_l$
\begin{eqnarray}
\mu_l = \D{E}[\tilde{u_l}] &  \approx & \int_{\Omega} \left (
\sum_{\alpha \leq r} u_{\alpha,l}
 H_{\alpha}(\vek{\xi}) \right ) \, \mathrm{d} \D{P} (\vek{\xi}) =
 u_{0,l}, \label{eq:ar01} \\
\sigma^{\mathrm{STD}}_l = \sqrt{\D{E}[(\tilde{u}_l -
\D{E}[\tilde{u}_l])^2]} & \approx & \sqrt{\left ( \sum_{0 < \alpha
\leq r} \gamma_{\alpha} \, u_{\alpha,l}^2 \right )}.
\label{eq:ar02}
\end{eqnarray}

\begin{figure}[h!]
\centering
\begin{tabular}{ccc}
\includegraphics[keepaspectratio,width=4.5cm]{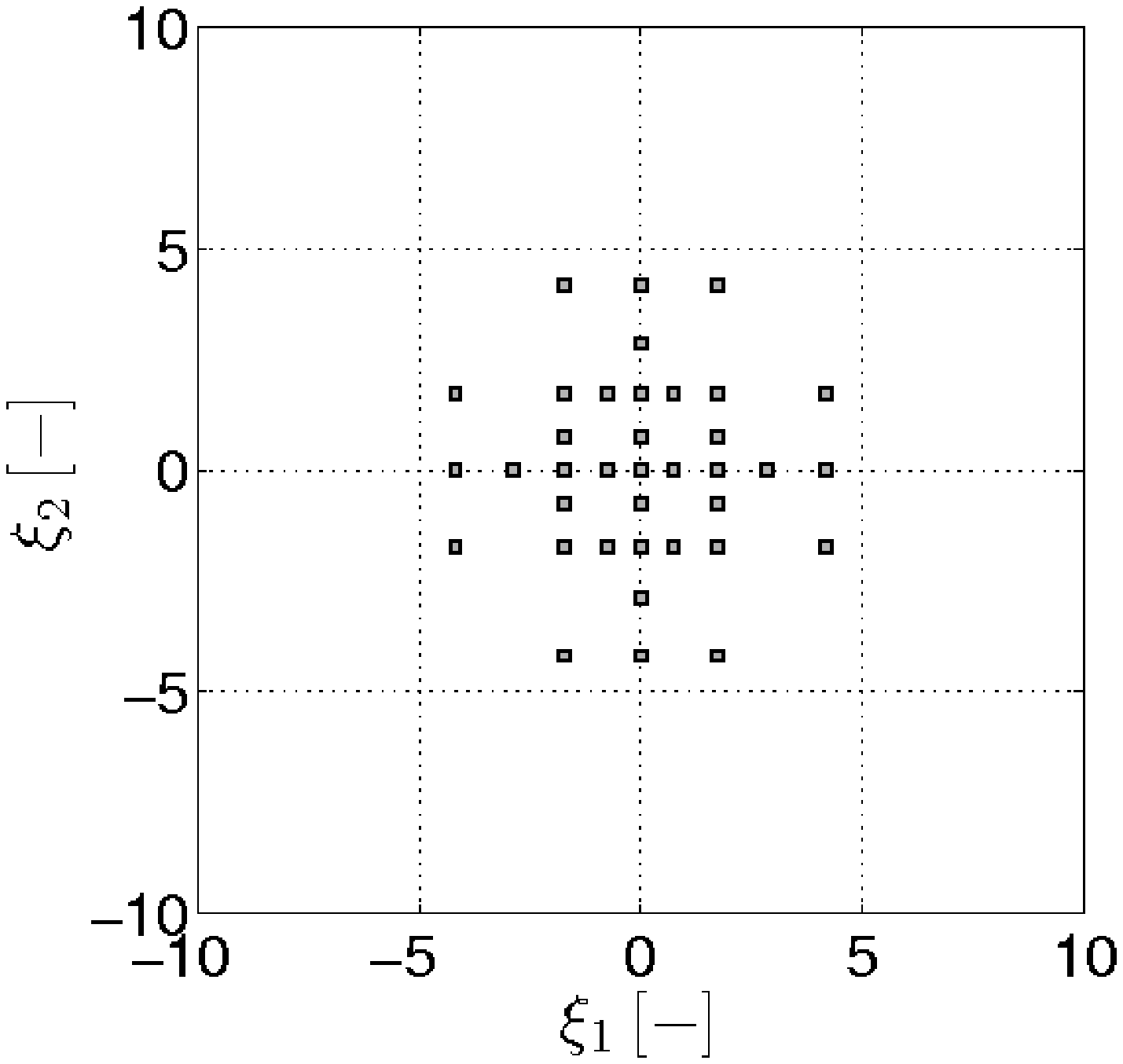} &
\includegraphics[keepaspectratio,width=4.5cm]{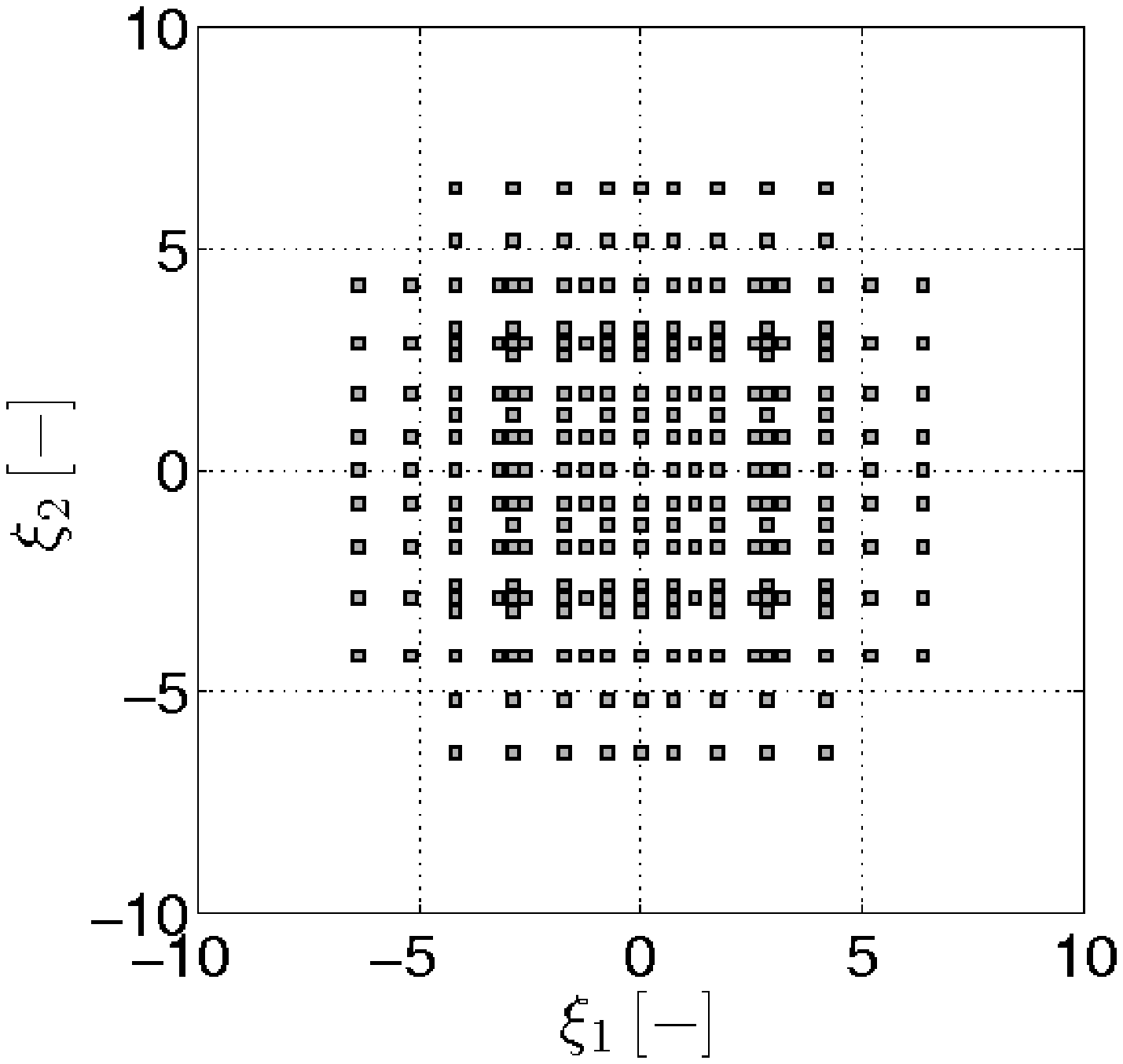} &
\includegraphics[keepaspectratio,width=4.5cm]{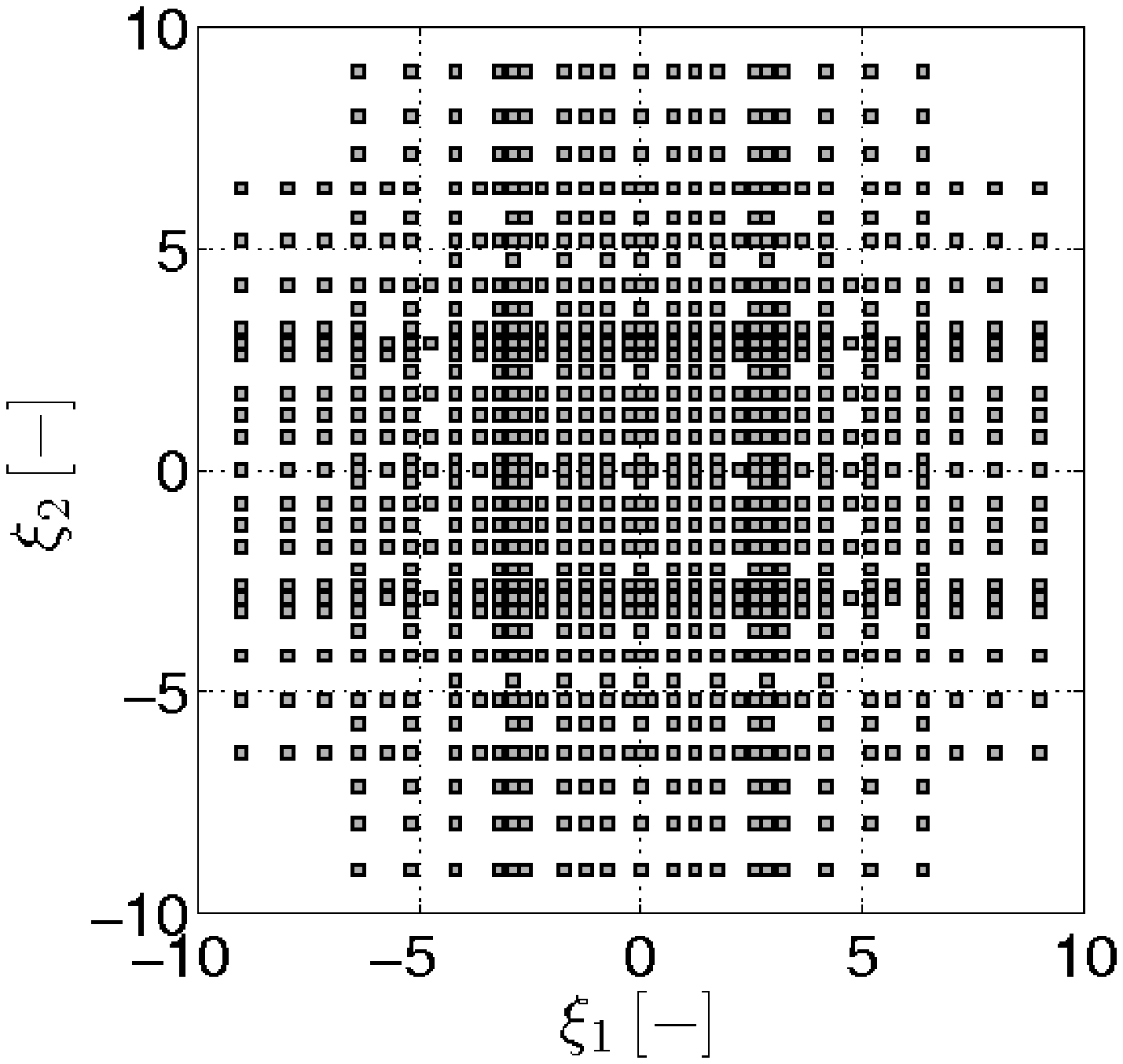} \\
(a) & (b) & (c) \\
\end{tabular}
\caption{Kronrod-Patterson quadrature rule for two Gaussian random
variables and different number of collocation points: (a) $i=37$;
(b) $i=261$; (c) $i=921$} \label{fig:sg}
\end{figure}

According to Eq.~(\ref{eq:sc1}), the computational effort of SC
method is given by the effort needed for realisation of $i$
deterministic simulations. Particular number of simulations $i$
and their spatial positions follow from choice of a desired
accuracy. Here we employ version of the Smolyak quadrature rule,
in particular nested Kronrod-Patterson version,
see~\cite{Heiss:2008}. This methodology produces significantly
less collocation points than other quadrature rules,
see~\cite{Xiu:2009}. For a detailed mathematical formulation and
corresponding numerical technique of several sparse quadratures
construction, we refer the interested reader
to~\cite{Babuska:2004,Babuska:2007,Heiss:2008,Xiu:2005}. For an
illustration, Kronrod-Patterson sparse grids for two Gaussian
random variables and different accuracy levels (number of points)
are plotted in Fig.~\ref{fig:sg}.

\section{Example: Plasticity problem at a material point}
\label{sec:exam}

\renewcommand{\arraystretch}{1.3}
%%%%%%%%%%%%%%%%%%%%%%%%%%%%%%%%%%%%%%%%%%%%%%%%%%%%%%%%%%%%%%%%%
\begin{table}[b!]
\begin{center}
\begin{tabular}{lclll}
 Symbol & Type of variable & Value & Mean ($\mu_q$) & Standard deviation ($\sigma^{\mathrm{STD}}_q$) \\
\hline
$E$  & lognormal RV & - & $210\cdot 10^{9}$ & $21\cdot 10^{9}$ \\
$\nu$ & constant & $0.3$ & - & - \\
$\sigma_{\mathrm{y,0}}$ & constant & $235\cdot 10^{6}$ & - & - \\
$H$ & constant & $2.1\cdot 10^{9}$ & - & - \\
\end{tabular}
\caption{Input material parameters of numerical study}
\label{tab:matpar1}
\end{center}
\end{table}
%%%%%%%%%%%%%%%%%%%%%%%%%%%%%%%%%%%%%%%%%%%%%%%%%%%%%%%%%%%%%%%%%

So as to demonstrate the described methodology on elasto-plastic
problem, we employ a numerical simulation of uniaxial tensile test
at a material point. For a sake of clarity, we introduce a simple
elasto-plastic model with four material parameters listed in
Tab.~\ref{tab:matpar1} and we consider only Young's modulus
$E\,\mathrm{[Pa]}$ to be uncertain. With respect to its physical
meaning, we describe it by a lognormally distributed RV. Remaining
three input parameters, namely Poisson's ratio
$\nu\,\mathrm{[-]}$, initial yield strength
$\sigma_{\mathrm{y,0}}\,\mathrm{[Pa]}$ and hardening parameter
$H\,\mathrm{[Pa]}$ are assumed to be a constant. The solution of
the elasto-plastic problem involves a discretization into $80$
uniform time steps $T$.

\begin{figure}[h!]
\centering
\begin{tabular}{c}
\includegraphics[keepaspectratio,width=8cm]{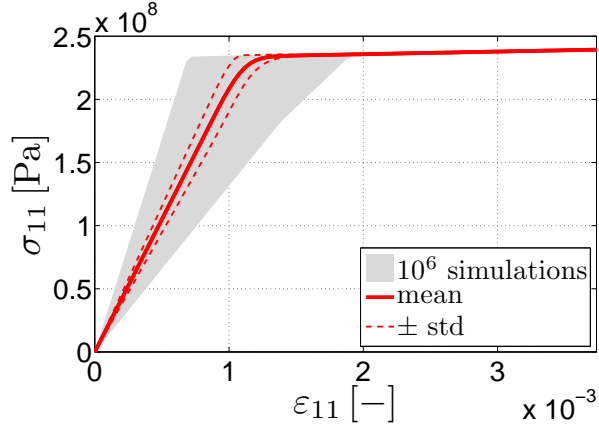} \\
\end{tabular}
\caption{Reference MC simulations corresponding to material
parameters in Tab.~\ref{tab:matpar1}, evolution of
$\sigma_{11}\,\mathrm{[Pa]}$ as function of
$\varepsilon_{11}\,\mathrm{[-]}$} \label{fig:rs1}
\end{figure}

First of all, we investigate the accuracy of SC-based strategy in
predicting mean, standard deviation and $0.01-$quantile of model
response $\sigma_{11}\,\mathrm{[Pa]} $\footnote{Other response
components are equal to zero.}. The quality of a PC-based
surrogate depends on the number $i$ of collocation points and on
the degree of polynomials $p$ used in Eq.~(\ref{eq:Tapprox}). In
order to assess the accuracy of surrogate models constructed for
different number of points $i$ and polynomial degree $p$, we
compare the obtained predictions with a reference solution
computed by MC method using $10^{6}$ samples, see
Fig.~\ref{fig:rs1}. To quantify the difference between the
resulting predictions of mean $\mu$, standard deviation
$\sigma^{\mathrm{STD}}$ and quantile $Q^{0.01}$ of model response
$\sigma_{11}\,\mathrm{[Pa]}$, we define following relative error:
\begin{eqnarray}
e(v)&=&\frac{\|v - v_{\mathrm{MC}}
\|_{l^2([0,T]\times\Omega)}}{\|v_{\mathrm{MC}}
\|_{l^2([0,T]\times\Omega)}}, \label{eq:L21}
%e(\mu)&=&\frac{\|\mu_{N} - \mu_{\mathrm{MC}}
%\|_{l^2([0,T]\times\Omega)}}{\|\mu_{\mathrm{MC}}
%\|_{l^2([0,T]\times\Omega)}}, \label{eq:L21}\\
%e(\sigma^{\mathrm{STD}})&=&\frac{\|\sigma^{\mathrm{STD}}_{N} -
%\sigma^{\mathrm{STD}}_{\mathrm{MC}}
%\|_{l^2([0,T]\times\Omega)}}{\|\sigma^{\mathrm{STD}}_{\mathrm{MC}}
%\|_{l^2([0,T]\times\Omega)}}, \label{eq:L22}
\end{eqnarray}
\begin{figure}[h!]
\centering
\begin{tabular}{ccc}
\includegraphics[keepaspectratio,width=4.5cm]{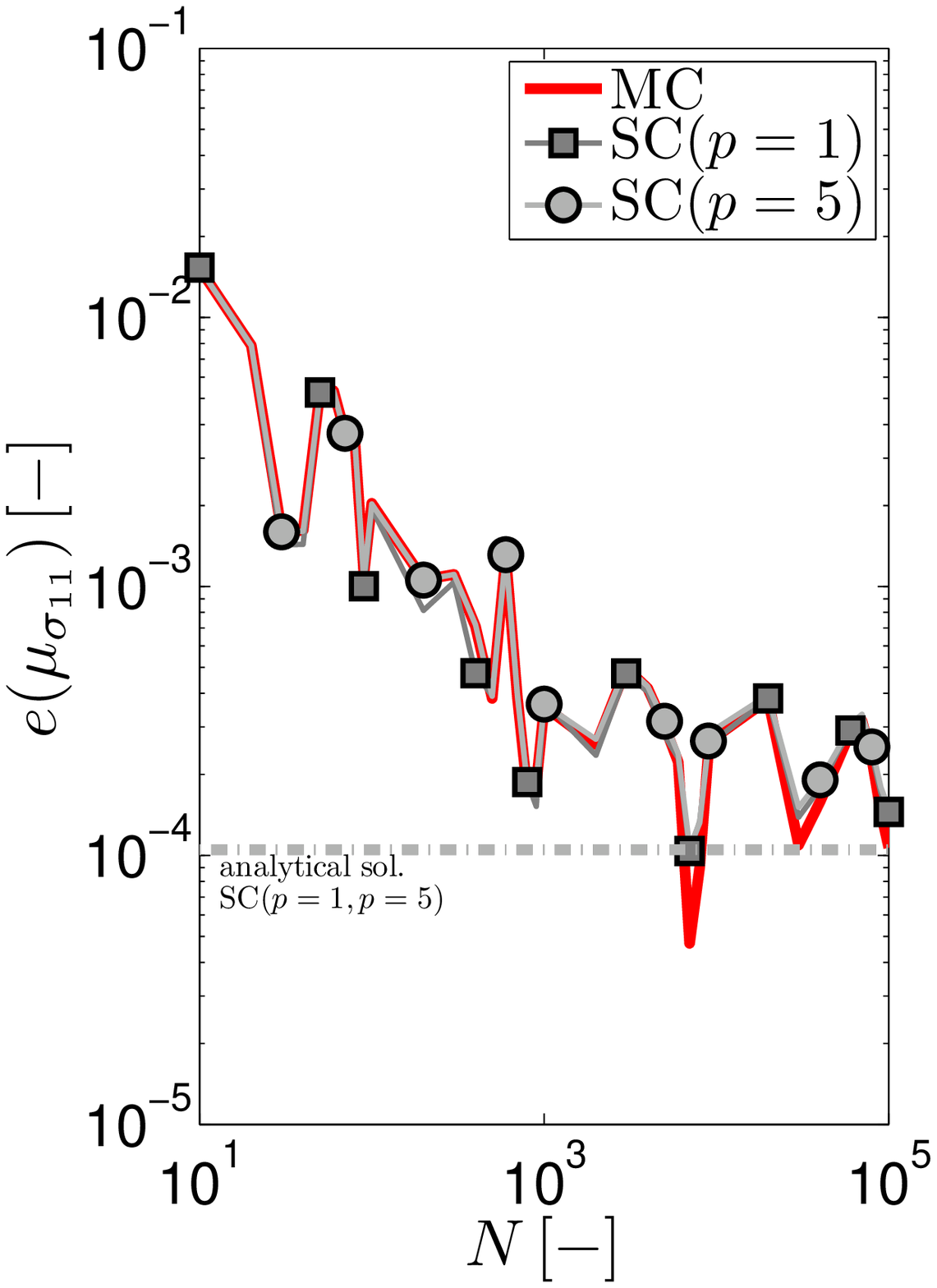} &
\includegraphics[keepaspectratio,width=4.5cm]{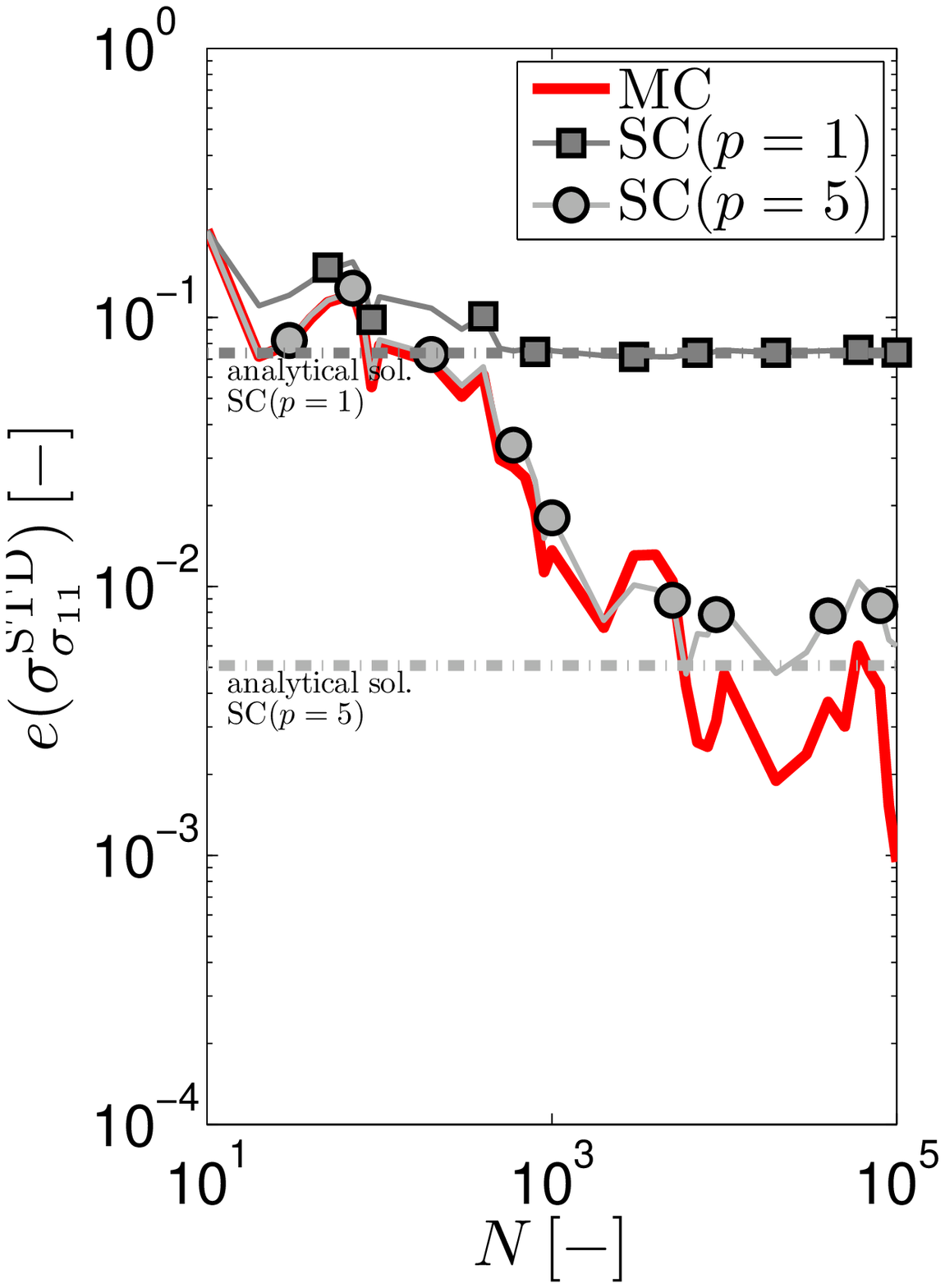} &
\includegraphics[keepaspectratio,width=4.5cm]{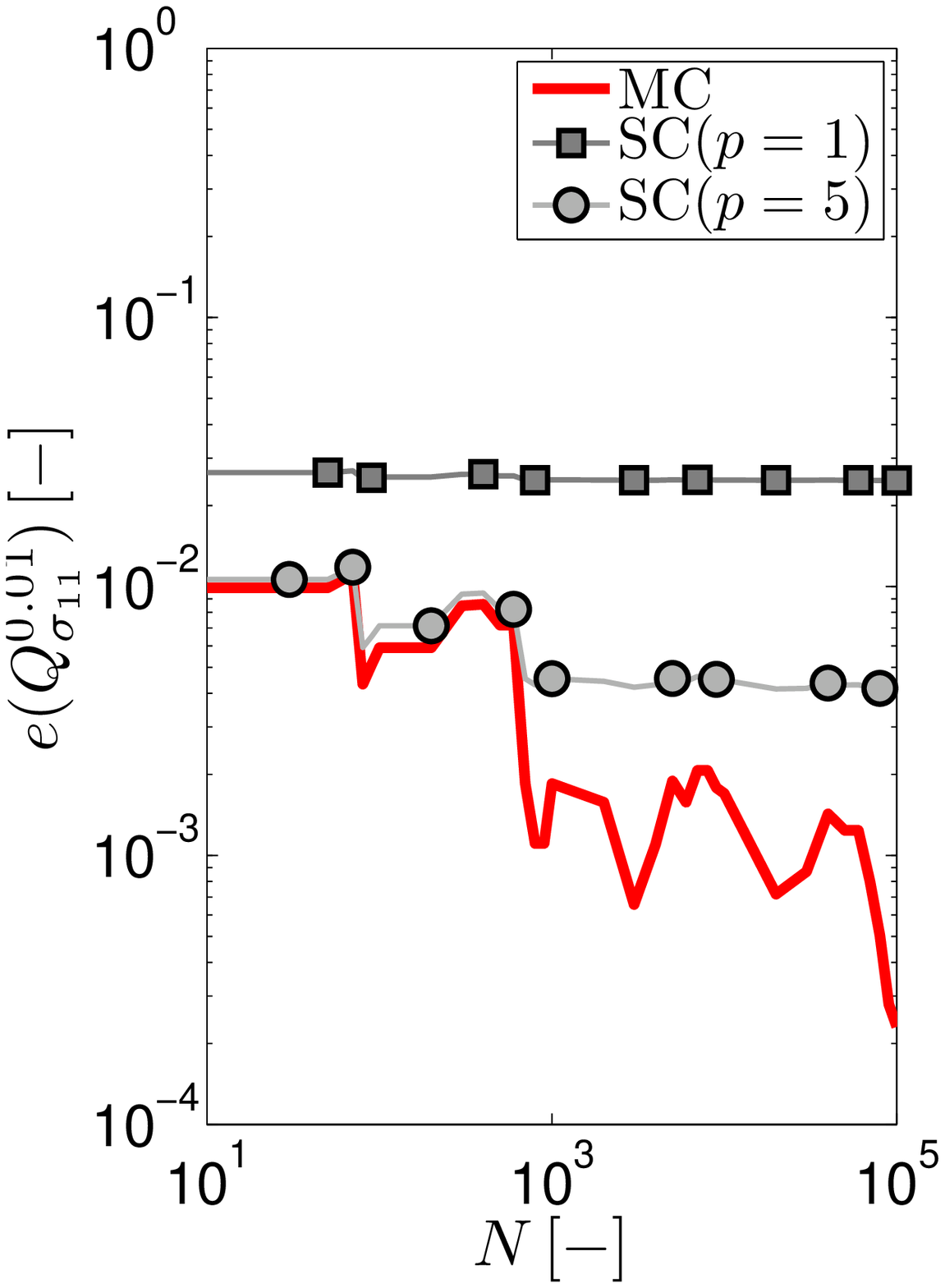}\\
(a) & (b) & (c) \\
\includegraphics[keepaspectratio,width=4.5cm]{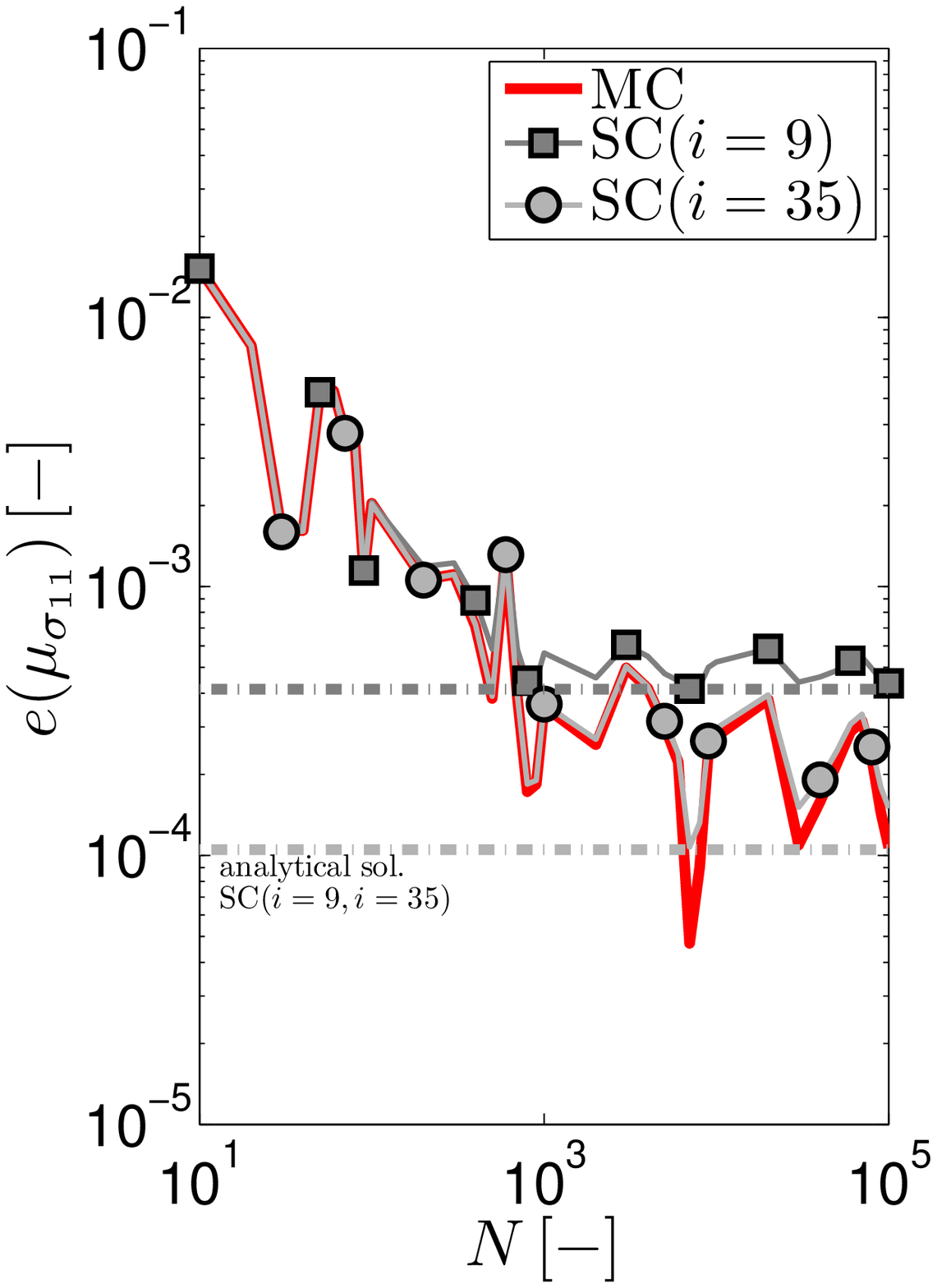} &
\includegraphics[keepaspectratio,width=4.5cm]{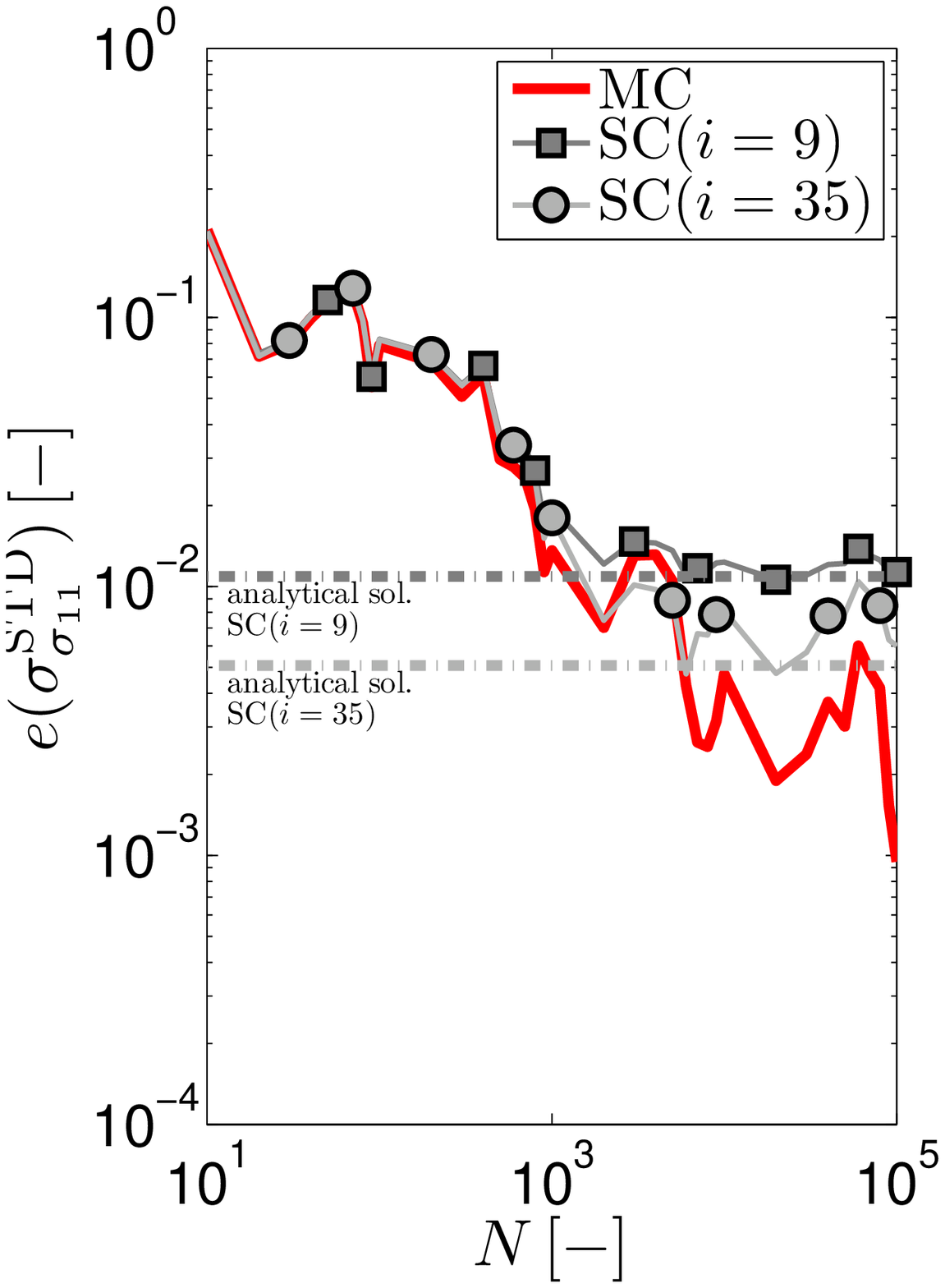} &
\includegraphics[keepaspectratio,width=4.5cm]{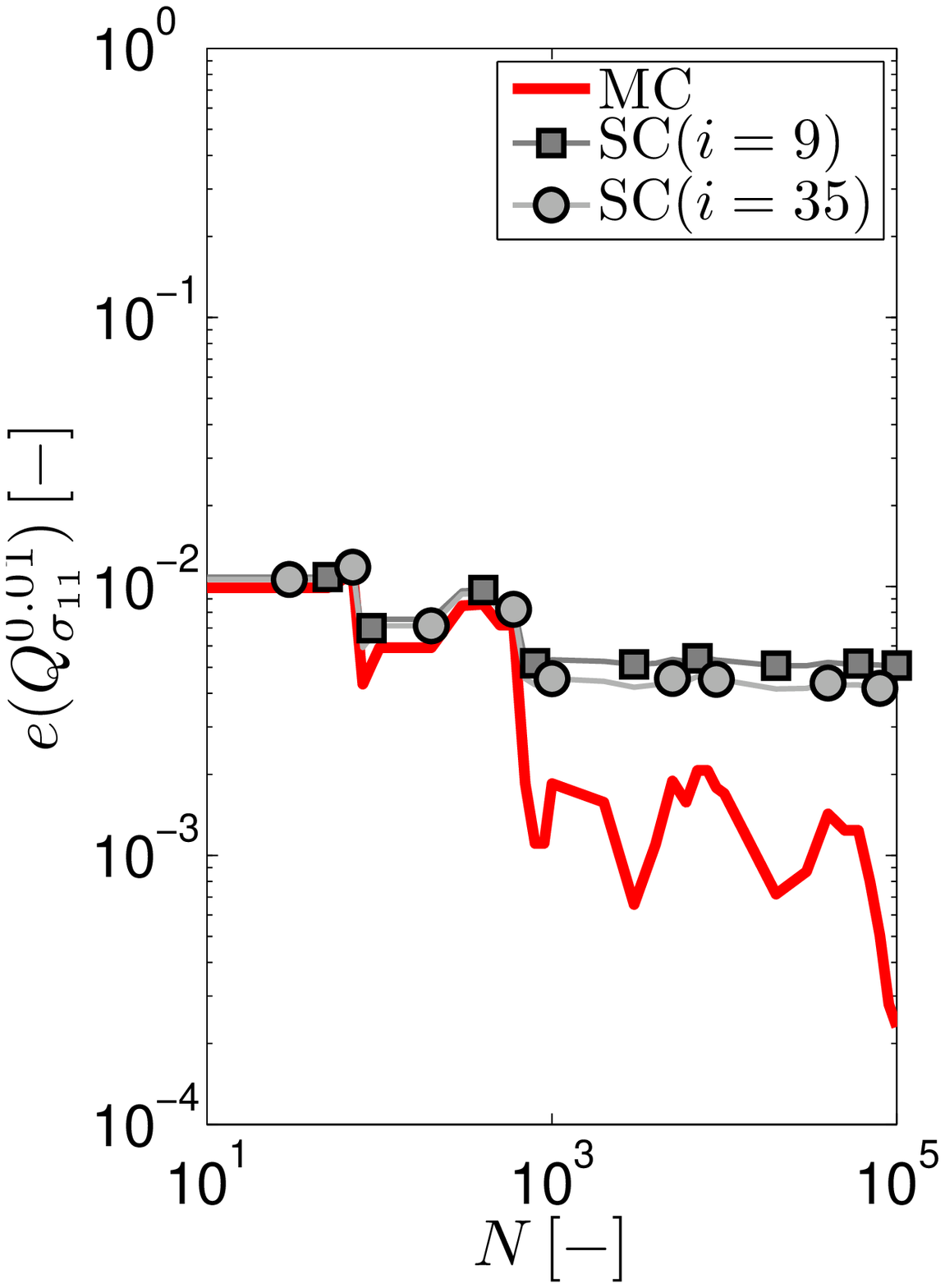}\\
(d) & (e) & (f) \\
\end{tabular}
%\caption{Error analysis of mean $\mu\,\mathrm{[-]}$ (a) and (b) and
%  standard deviation $\sigma^{\mathrm{STD}}\,\mathrm{[-]}$ (c) and (d)
%  for fixed number of sparse grid points $(i=35)$ (a) and (c) and for
%  fixed degree of polynomial chaos expansion $(p=5)$ (b) and (d)}
\caption{Error analysis of (a) mean $\mu\,\mathrm{[-]}$, (b) standard
  deviation $\sigma^{\mathrm{STD}}\,\mathrm{[-]}$ and (c) quantile
  $Q^{0.01}\,\mathrm{[-]}$ for fixed number of sparse grid points
  $(i=35)$; (d) Error analysis of mean $\mu\,\mathrm{[-]}$, (e)
  standard deviation $\sigma^{\mathrm{STD}}\,\mathrm{[-]}$ and (f)
  quantile $Q^{0.01}\,\mathrm{[-]}$ for fixed degree of PCE $(p=5)$}
\label{fig:err1}
\end{figure}
where $v$ stands for an investigated prediction of mean, standard
deviation or quantile, $v_{\mathrm{MC}}$ denotes the reference MC
solution and $l^2$ is the Euclidean norm. The prediction based on
a surrogate model is calculated using either MC sampling method or
-- if possible -- using analytical relations, which are
independent on the sampling procedure, see Eqs.~(\ref{eq:ar01})
and (\ref{eq:ar02}). Figs.~\ref{fig:err1}(a)-(f) display the
evolution of error in sampling-based predictions (solid lines) of
mean $\mu$, standard deviation $\sigma^{\mathrm{STD}}$ and
quantile $Q^{0.01}$ along with the number of MC samples used for
their computation. Particularly, Figs.~\ref{fig:err1}(a)-(c)
compare the results for different degree of PC expansion ($p=1$
and $p=5$) assembled with the same number of collocation points
$i=35$, while Figs.~\ref{fig:err1}(d)-(f) display the error in
predictions for different numbers of collocation points ($i=9$ and
$i=35$) and fixed polynomial degree $p=5$. All the sampling-based
predictions were obtained using the same MC samples and thus, the
differences among the curves depicted in
Figs.~\ref{fig:err1}(a)-(f) are produced solely by the inaccuracy
of the involved PCE. The sampling-based predictions are also
accompanied by the predictions obtained analytically from the
constructed PCE (dashed lines). Obviously, the predictions
computed by sampling of a chosen PCE converge towards the
predictions obtained from the same PCE analytically.
%
%\renewcommand{\arraystretch}{1.3}
%%%%%%%%%%%%%%%%%%%%%%%%%%%%%%%%%%%%%%%%%%%%%%%%%%%%%%%%%%%%%%%%%%
%\begin{table}[b!]
%\begin{center}
%\begin{tabular}{c|cp{0.5cm}ccccc}
% & $e(\mu_{\sigma_{11}})$ & &  $e(\sigma^{\mathrm{STD}}_{\sigma_{11}})$ & & & & \\
%$i$ $|$ $p$ & - & & 1 & 2 & 3 & 4 & 5 \\
%\cline{1-2} \cline{4-8}
%9 & $4.1\cdot 10^{-4}$ &   & $7.5\cdot 10^{-2}$ & $2.6\cdot 10^{-2}$ & $1.4\cdot 10^{-2}$ & $1.1\cdot 10^{-2}$ & $1.1\cdot 10^{-2}$ \\
%17 & $3.9\cdot 10^{-4}$ &  & $7.4\cdot 10^{-2}$ & $2.5\cdot 10^{-2}$ & $1.4\cdot 10^{-2}$ & $1.0\cdot 10^{-2}$ & $9.8\cdot 10^{-3}$ \\
%19 & $3.4\cdot 10^{-4}$ &  & $7.4\cdot 10^{-2}$ & $2.4\cdot 10^{-2}$ & $1.2\cdot 10^{-2}$ & $8.6\cdot 10^{-3}$ & $7.4\cdot 10^{-3}$ \\
%33 & $1.7\cdot 10^{-4}$ &  & $7.4\cdot 10^{-2}$ & $2.4\cdot 10^{-2}$ & $1.1\cdot 10^{-2}$ & $7.1\cdot 10^{-3}$ & $5.6\cdot 10^{-3}$ \\
%35 & $1.0\cdot 10^{-4}$ &  & $7.4\cdot 10^{-2}$ & $2.4\cdot 10^{-2}$ & $1.1\cdot 10^{-2}$ & $6.8\cdot 10^{-3}$ & $5.1\cdot 10^{-3}$ \\
%\cline{1-2} \cline{4-8}
%\end{tabular}
%\caption{Errors of surrogate models calculated using analytical
%relations} \label{tab:err1}
%\end{center}
%\end{table}
\renewcommand{\arraystretch}{1.3}
%%%%%%%%%%%%%%%%%%%%%%%%%%%%%%%%%%%%%%%%%%%%%%%%%%%%%%%%%%%%%%%%%
\begin{table}[b!]
\begin{center}
\begin{tabular}{c|c|ccccc}
 & $p$ $/$ $i$ & 9 & 17 & 19 & 33 & 35 \\
 \hline
 $e(\mu_{\sigma_{11}})$ & - & $4.1\cdot 10^{-4}$ & $3.9\cdot 10^{-4}$ & $3.4\cdot 10^{-4}$ & $1.7\cdot 10^{-4}$ & $1.0\cdot 10^{-4}$ \\
 \hline
 $e(\sigma^{\mathrm{STD}}_{\sigma_{11}})$ & 1 & $7.5\cdot 10^{-2}$ & $7.4\cdot 10^{-2}$ & $7.4\cdot 10^{-2}$ & $7.4\cdot 10^{-2}$ & $7.4\cdot 10^{-2}$ \\
  & 3 & $1.4\cdot 10^{-2}$ & $1.4\cdot 10^{-2}$ & $1.2\cdot 10^{-2}$ & $1.1\cdot 10^{-2}$ & $1.1\cdot 10^{-2}$ \\
  & 5 & $1.1\cdot 10^{-2}$ & $9.8\cdot 10^{-3}$ & $7.4\cdot 10^{-3}$ & $5.6\cdot 10^{-3}$ & $5.1\cdot 10^{-3}$ \\
 \hline
 $e(Q^{0.01}_{\sigma_{11}})$ & 1 & $2.5\cdot 10^{-2}$ & $2.5\cdot 10^{-2}$ & $2.5\cdot 10^{-2}$ & $2.5\cdot 10^{-2}$ & $2.5\cdot 10^{-2}$ \\
  & 3 & $5.1\cdot 10^{-3}$ & $4.9\cdot 10^{-3}$ & $4.7\cdot 10^{-3}$ & $4.6\cdot 10^{-3}$ & $4.6\cdot 10^{-3}$ \\
  & 5 & $5.1\cdot 10^{-3}$ & $4.9\cdot 10^{-3}$ & $4.4\cdot 10^{-3}$ & $4.2\cdot 10^{-3}$ & $4.1\cdot 10^{-3}$ \\
 \hline
\end{tabular}
\caption{Errors of surrogate models calculated for different
degree of PC expansion $p$ and number of collocation points $i$}
\label{tab:err1}
\end{center}
\end{table}
We would like to point out Fig.~\ref{fig:err1}(e), where all the
sampling-based predictions have comparable error for lower number
of samples. For approximately $10^3$ samples, the error of the
sampling-based predictions reaches the error of analytical
prediction and with more samples PCE-based predictions remain
almost unchanged while the error of the full model-based MC
sampling continues decreasing. In other words, difference between
curves in the left part of the graph clearly refers to the
sampling error, while the right part of the graph reveals the
error of the PCE-based surrogates.  Same phenomenon may be
observed in Fig.~\ref{fig:err1}(b), but in Fig.~\ref{fig:err1}(a)
is the error of surrogates negligible. We may conclude that with
enough collocation points already the degree $p=1$ is sufficient
for predicting mean of a model response, while
Fig.~\ref{fig:err1}(d) shows that high polynomial degree cannot
compensate low number of collocation points. Contrarily,
Figs.~\ref{fig:err1}(b)-(c) and~\ref{fig:err1}(e)-(f) indicate
that for predictions of higher statistical moments or quantiles,
higher order of polynomial degree becomes more important than the
employed number of collocation points.

Comparison of predictions for other values of polynomial degree $p$
and number of collocation points $i$ is listed in Tab.~\ref{tab:err1}.
Predictions of mean $\mu$ and standard deviation
$\sigma^{\mathrm{STD}}$ are computed from surrogates analytically.
From the relationship in Eqs.~(\ref{eq:ar01}) and (\ref{eq:ar02}) it
is clear that mean of the response is fully described by the constant
term of polynomials and higher degrees are relevant only for
prediction of standard deviation.  Number of collocation points is on
the other hand important for predicting both the statistical moments.
The results also confirm that predictions of standard deviation
$\sigma^{\mathrm{STD}}$ or quantile $Q^{0.01}$ are more sensitive to
polynomial degree than to the number of collocation points.

Furthermore, the coefficient of determination $R^{2}$ is utilised
to indicate the overall accuracy of PC-based surrogate models. It
provides a measure of how well is the reference MC solution
reproduced by a surrogate model in terms of the data variance
explained by the surrogate relative to the total data variance,
see~\cite{Steel:1960}. Hence, the values of $R^2$ lie between $0$
and $1$, and perfectly explained data variation is denoted by $1$.
The definition of the coefficient of determination $R^{2}$ is for
our purpose expressed as
\begin{equation}
R^{2} = 1 -
\frac{\|\sigma_{11,\mathrm{MC}}-\sigma_{11,\mathrm{PC}}\|^{2}_{l^2(\Omega)}}{\|\sigma_{11,\mathrm{MC}}-\mu_{\sigma_{11},\mathrm{MC}}\|^{2}_{l^2(\Omega)}}.
\label{eq:R2}
\end{equation}

\begin{figure}[h!]
\centering
\begin{tabular}{cc}
\includegraphics[keepaspectratio,width=6.5cm]{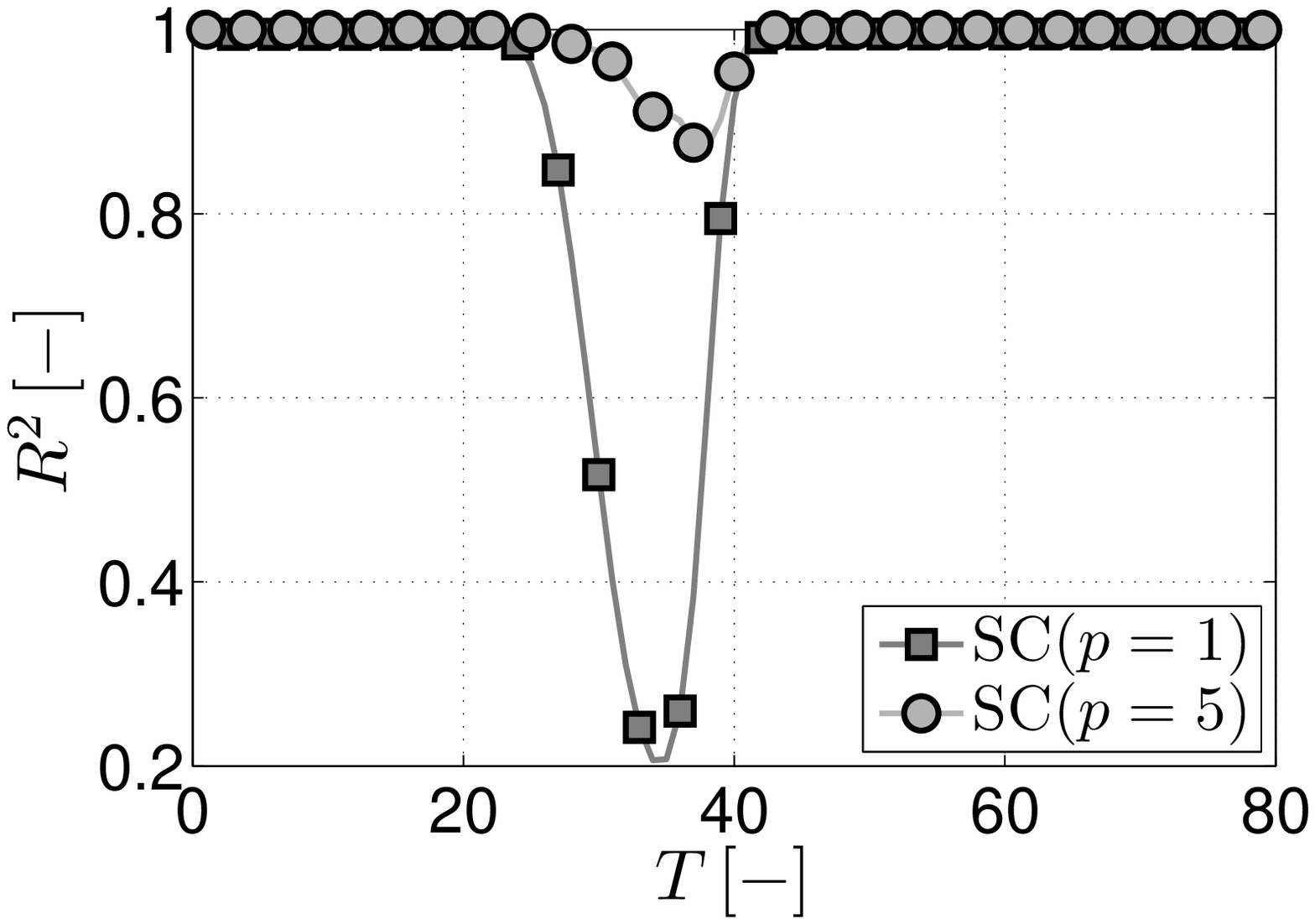} &
\includegraphics[keepaspectratio,width=6.5cm]{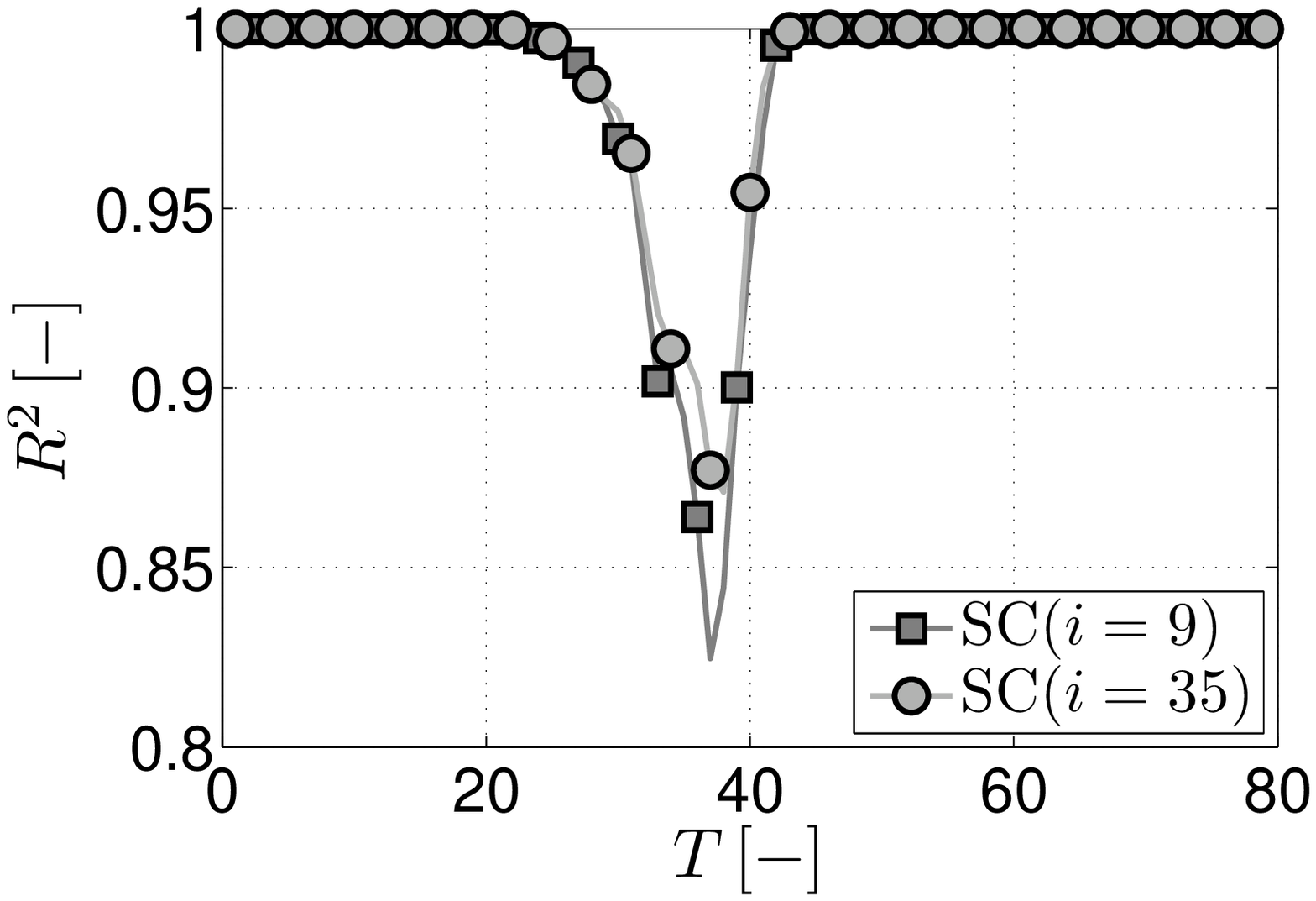}\\
(a) & (b) \\
\end{tabular}
\caption{Evolution of $R^2$: (a) as a
  function of degree of PCE ($i=35$); (b) as a
  function of sparse grid points ($p=5$)} \label{fig:R21}
\end{figure}
\begin{figure}[t!]
\centering
\begin{tabular}{cc}
\includegraphics[keepaspectratio,width=6.5cm]{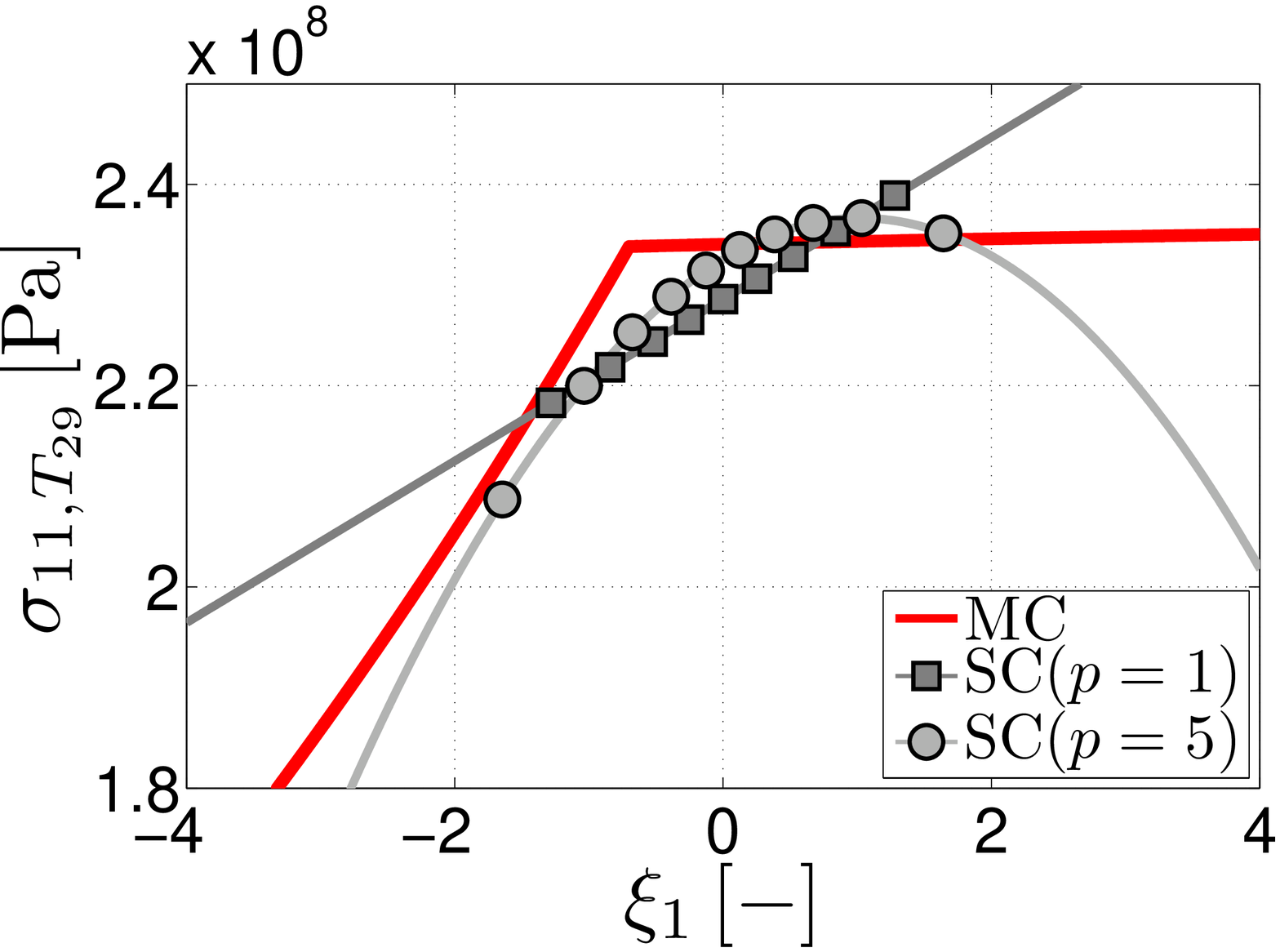} &
\includegraphics[keepaspectratio,width=6.5cm]{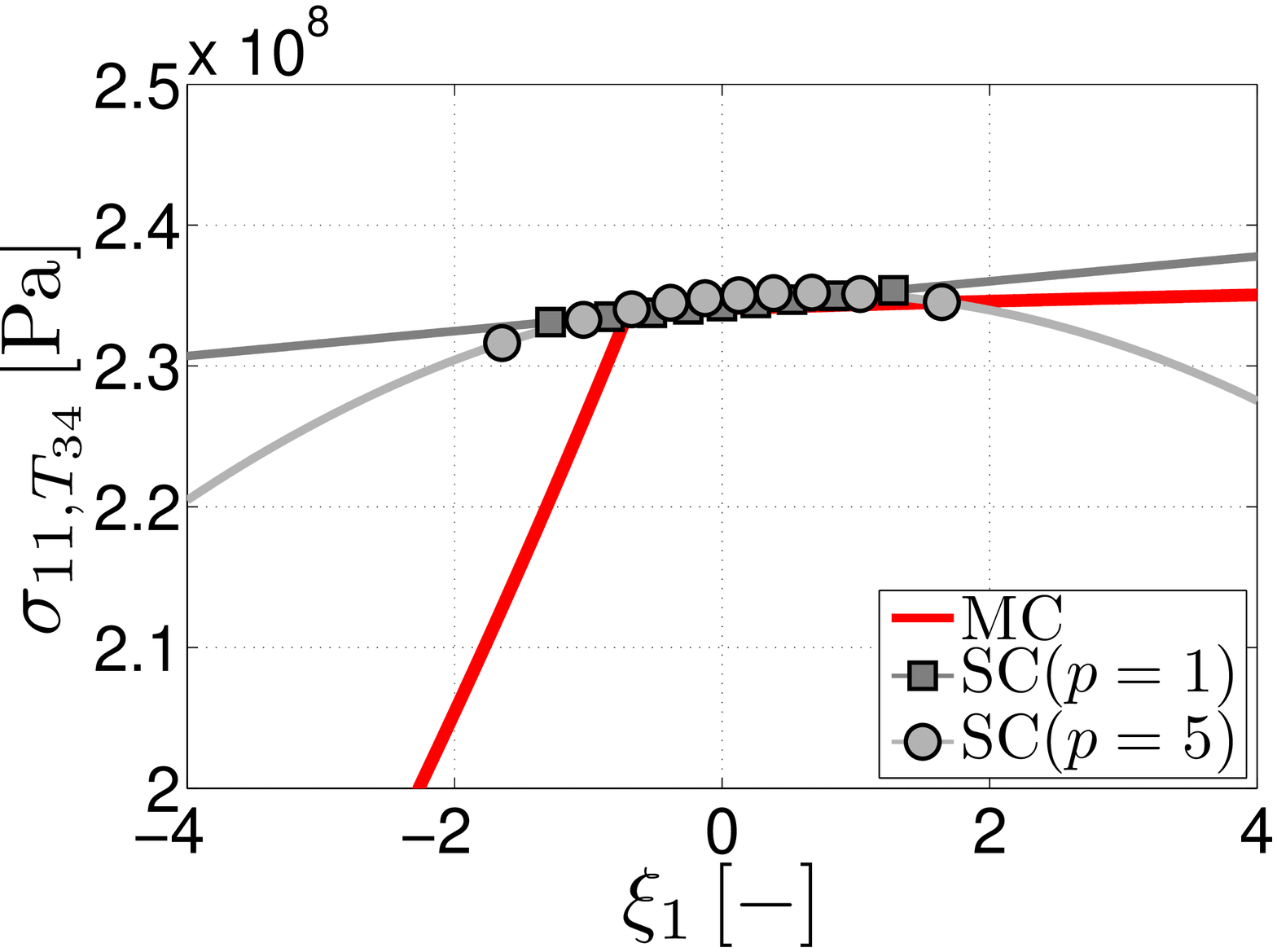}\\
(a) & (b) \\
\end{tabular}
\caption{Shape of critical PC expansions corresponding to (a)
$T=29$
  and (b) $T=34$.}
\label{fig:T30}
\end{figure}

Figs.~\ref{fig:R21}(a)-(b) display the evolution of the
coefficient of determination $R^2$ in all time steps of numerical
example. It is evident that the surrogate model is less accurate
when the elastic limit is reached and the model response is not
smooth. This phenomenon is displayed in Figs.~\ref{fig:T30}(a)-(b)
where two examples of PC expansions corresponding to time steps
$T=29$ and $T=34$, respectively, are plotted as functions of
random variable $\xi_{1}$ and compared with the reference MC
solution.  One can see that as a consequence of active yielding,
the model response $\sigma_{11}\,\mathrm{[Pa]}$ is not smooth, but
close to bilinear, which is difficult to be approximated by low
order polynomials. On the other hand, Fig~\ref{fig:R21}(b) points
out that once having sufficient polynomial degree, increasing the
number collocation points does not lead to significant improvement
of surrogate quality as documented also by results in
Tab.~\ref{tab:err1}.

\begin{figure}[t!]
\centering
\begin{tabular}{c}
\includegraphics[keepaspectratio,width=8.5cm]{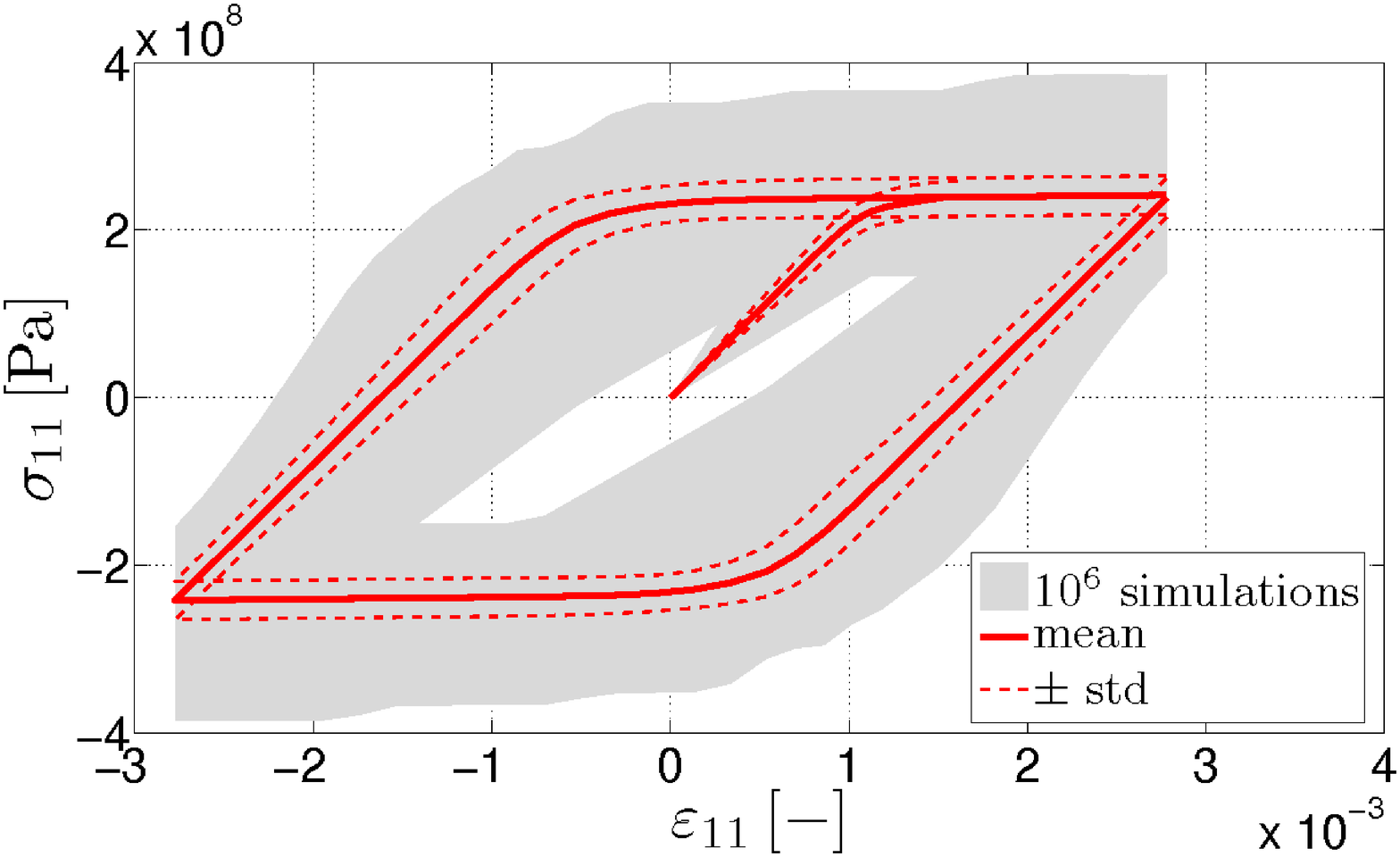} \\
\end{tabular}
\caption{Reference MC simulations corresponding to material
parameters in Tab.~\ref{tab:matpar2}, evolution of model response
$\sigma_{11}\,\mathrm{[Pa]}$ as function of
$\varepsilon_{11}\,\mathrm{[-]}$} \label{fig:rs2}
\end{figure}

Several interesting results have been derived within the scope of
numerical calculations with all uncertain parameters as an input.
We consider all the parameters to be lognormally distributed with
prescribed mean $\mu_q$ and standard deviation
$\sigma^{\mathrm{STD}}_q$ given in Tab.~\ref{tab:matpar2}.
\renewcommand{\arraystretch}{1.2}
%%%%%%%%%%%%%%%%%%%%%%%%%%%%%%%%%%%%%%%%%%%%%%%%%%%%%%%%%%%%%%%%%
\begin{table}[b]
\begin{center}
\begin{tabular}{lclll}
 Symbol &  Type of variable & Value & Mean ($\mu_q$) & Standard deviation ($\sigma^{\mathrm{STD}}_q$) \\
\hline
$E$   & lognormal RV & - & $210\cdot 10^{9}$ & $21\cdot 10^{9}$ \\
$\nu$   & lognormal RV & - & $0.3$ & $0.015$ \\
$\sigma_{\mathrm{y,0}}$  & lognormal RV & - & $235\cdot 10^{6}$ & $23.5\cdot 10^{6}$ \\
$H$  & lognormal RV & - & $21\cdot 10^{8}$ & $2.1\cdot 10^{8}$ \\
\end{tabular}
\caption{Input parameters of numerical study} \label{tab:matpar2}
\end{center}
\end{table}
%%%%%%%%%%%%%%%%%%%%%%%%%%%%%%%%%%%%%%%%%%%%%%%%%%%%%%%%%%%%%%%%%
%
In addition, the uniaxial tensile problem is extended by load/unload
cycle resulting in $300$ uniform time steps $T$ of numerical analysis.
Fig.~\ref{fig:rs2} presents the reference MC solution obtained for
$10^{6}$ samples.

\renewcommand{\arraystretch}{1.3}
%%%%%%%%%%%%%%%%%%%%%%%%%%%%%%%%%%%%%%%%%%%%%%%%%%%%%%%%%%%%%%%%%
\begin{table}[b!]
\begin{center}
\begin{tabular}{L{1.5cm}|c|C{2cm}C{2cm}C{2cm}C{2cm}}
 & $p$ $/$ $i$ & 201 & 3065 & 12057 & 20681  \\
 \hline
 $e(\mu_{\sigma_{11}})$ & - & $3.6\cdot 10^{-3}$ & $1.1\cdot 10^{-3}$ & $3.1\cdot 10^{-4}$ & $2.3\cdot 10^{-4}$ \\
 \hline
 $e(\sigma^{\mathrm{STD}}_{\sigma_{11}})$ & 1 & $7.9\cdot 10^{-2}$ & $7.8\cdot 10^{-2}$ & $7.8\cdot 10^{-2}$ & $7.8\cdot 10^{-2}$ \\
  & 5 & $2.3\cdot 10^{-2}$ & $6.4\cdot 10^{-3}$ & $3.7\cdot 10^{-3}$ & $3.8\cdot 10^{-3}$ \\
  & 15 & -- & -- & $2.2\cdot 10^{-3}$ & $1.6\cdot 10^{-3}$ \\
 \hline
 $e(Q^{0.01}_{\sigma_{11}})$ & 1 & $4.1\cdot 10^{-2}$ & $4.1\cdot 10^{-2}$ & $4.1\cdot 10^{-2}$ & $4.1\cdot 10^{-2}$ \\
  & 5 & $1.3\cdot 10^{-2}$ & $7.6\cdot 10^{-3}$ & $7.2\cdot 10^{-3}$ & $7.1\cdot 10^{-3}$ \\
  & 15 & -- & -- & $2.5\cdot 10^{-3}$ & $2.4\cdot 10^{-3}$ \\
 \hline
\end{tabular}
\caption{Errors of surrogate models calculated for different
degree of PC expansion $p$ and number of collocation points $i$}
\label{tab:err2}
\end{center}
\end{table}

\begin{figure}[h!]
\centering
\begin{tabular}{ccc}
\includegraphics[keepaspectratio,width=4.5cm]{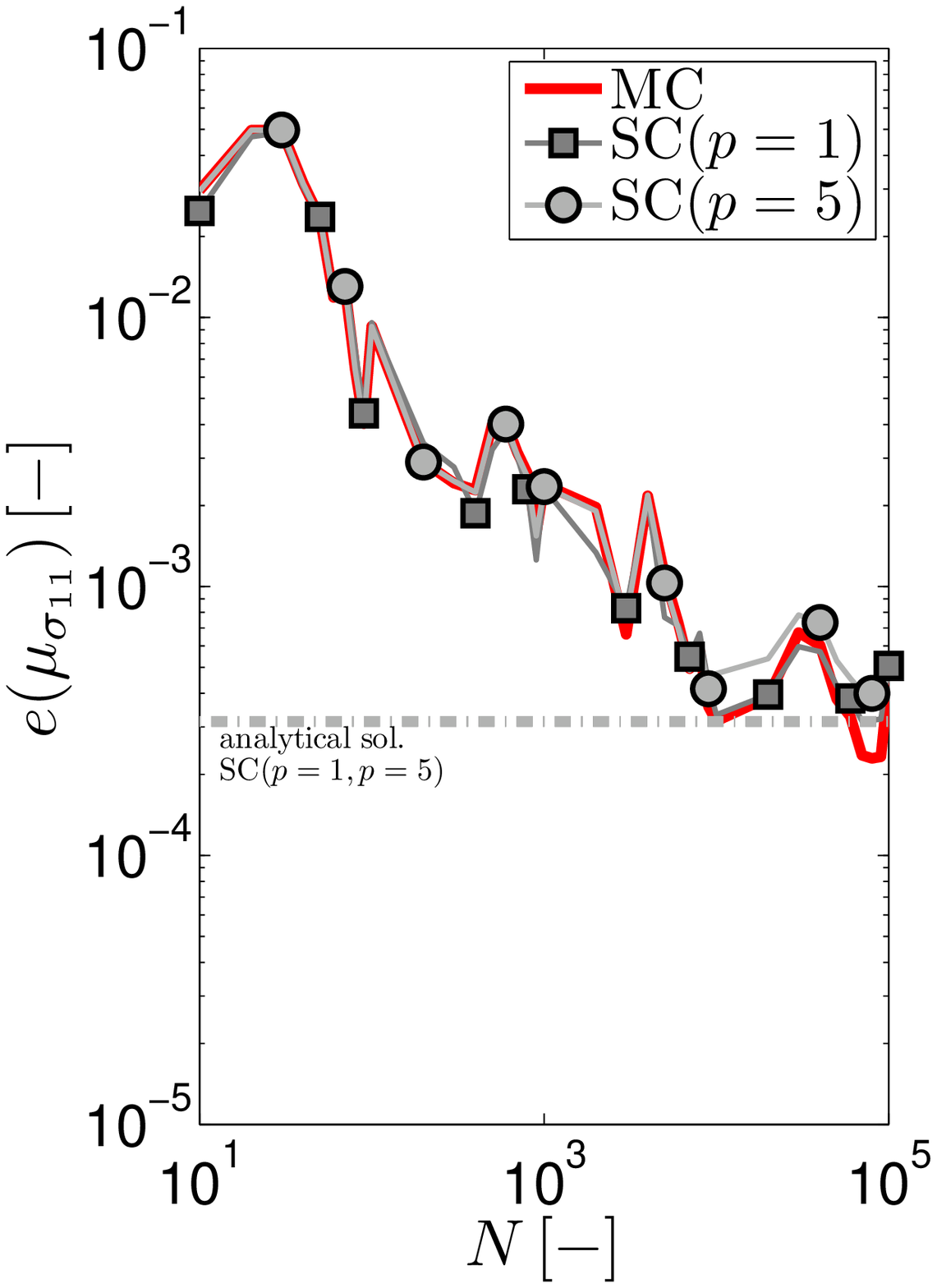} &
\includegraphics[keepaspectratio,width=4.5cm]{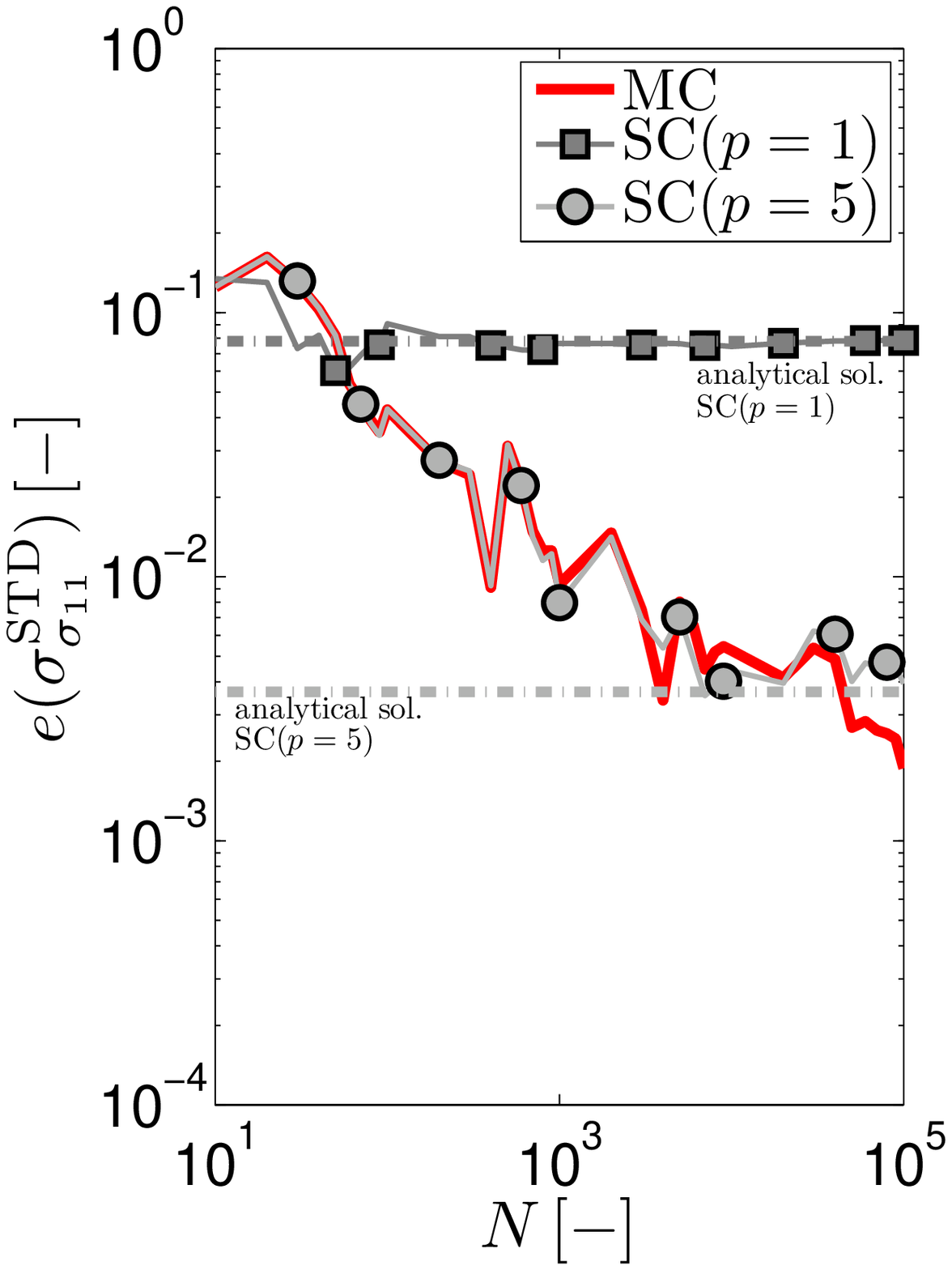} &
\includegraphics[keepaspectratio,width=4.5cm]{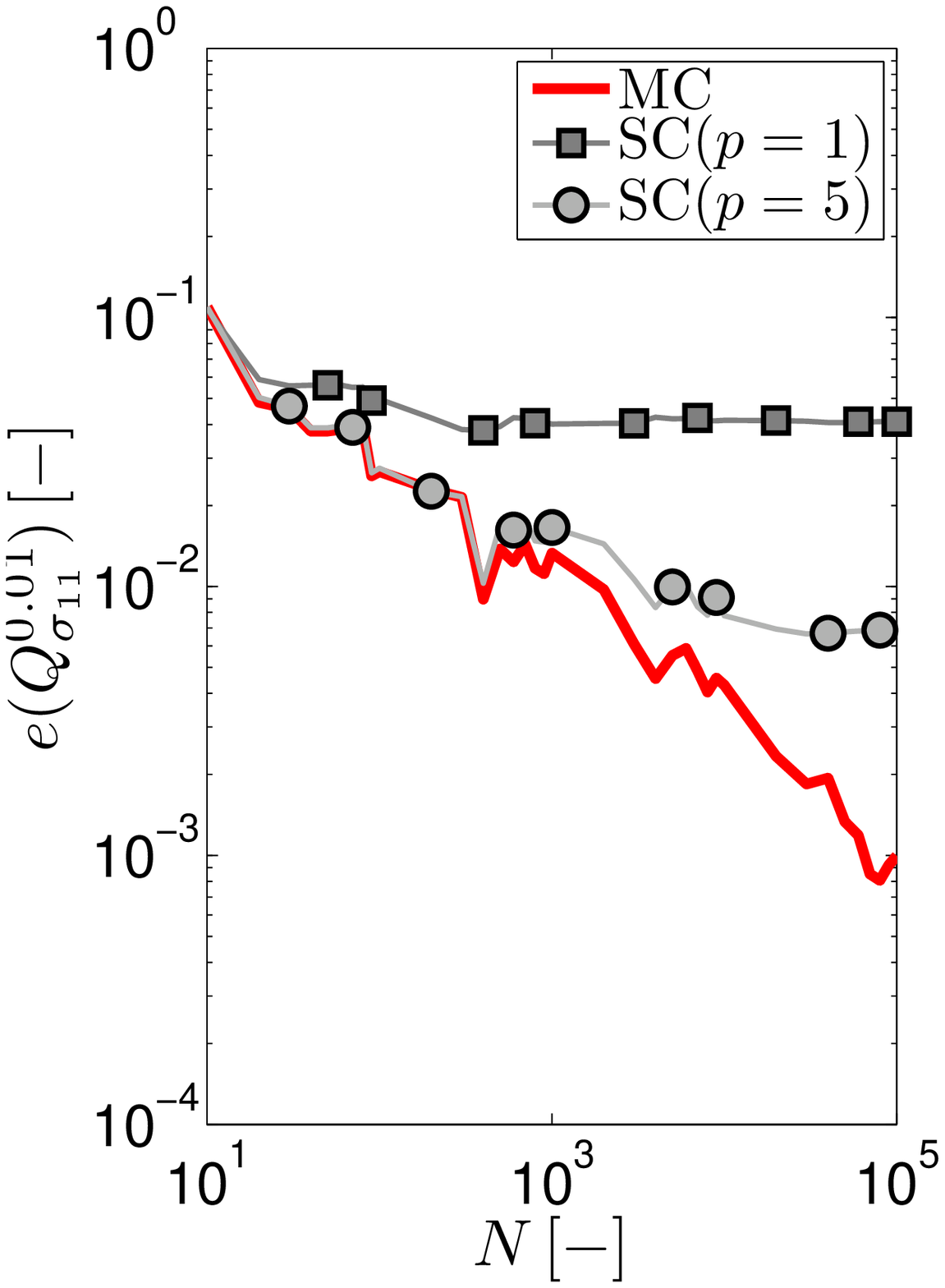} \\
(a) & (b) & (c) \\
\includegraphics[keepaspectratio,width=4.5cm]{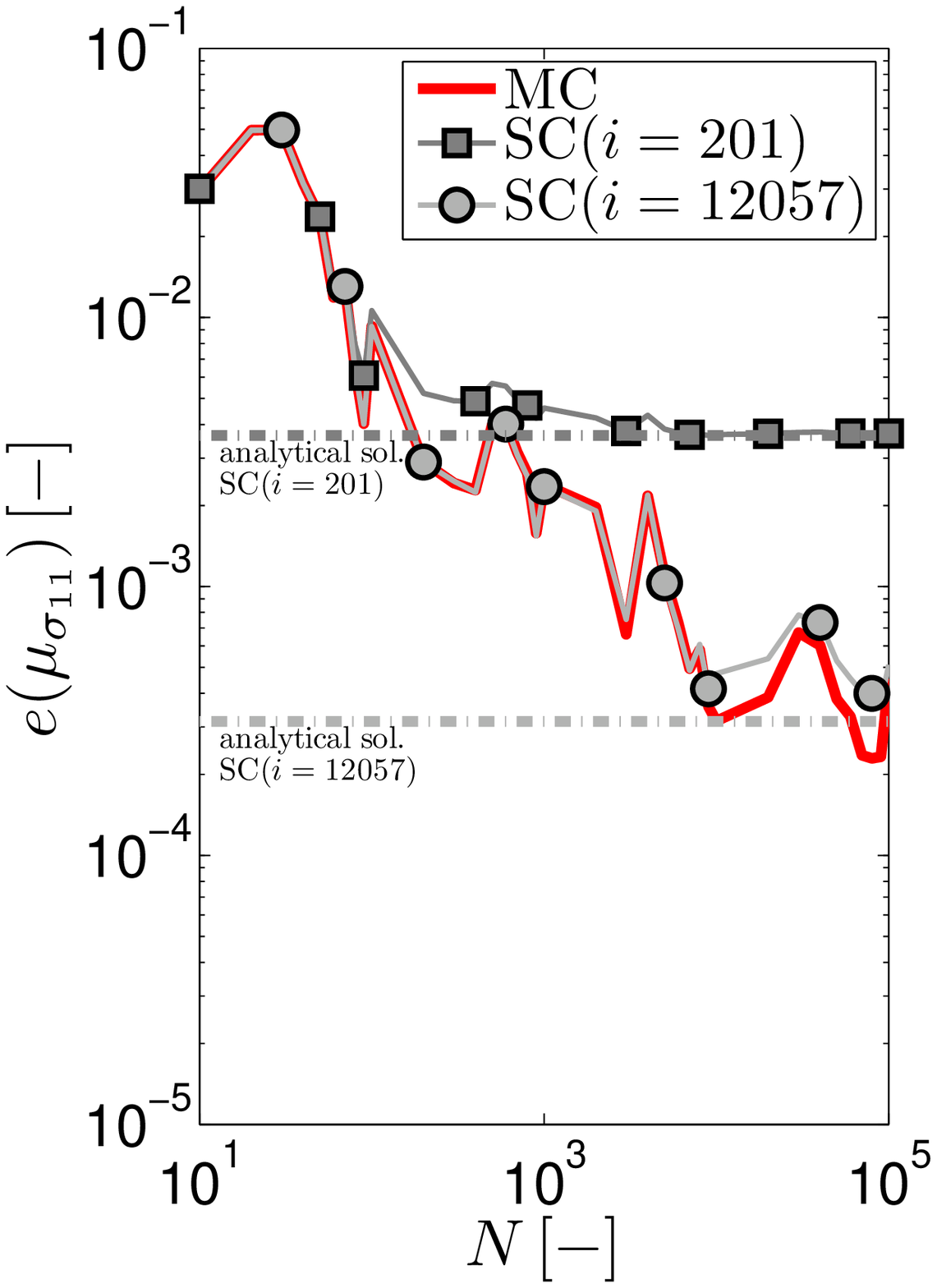} &
\includegraphics[keepaspectratio,width=4.5cm]{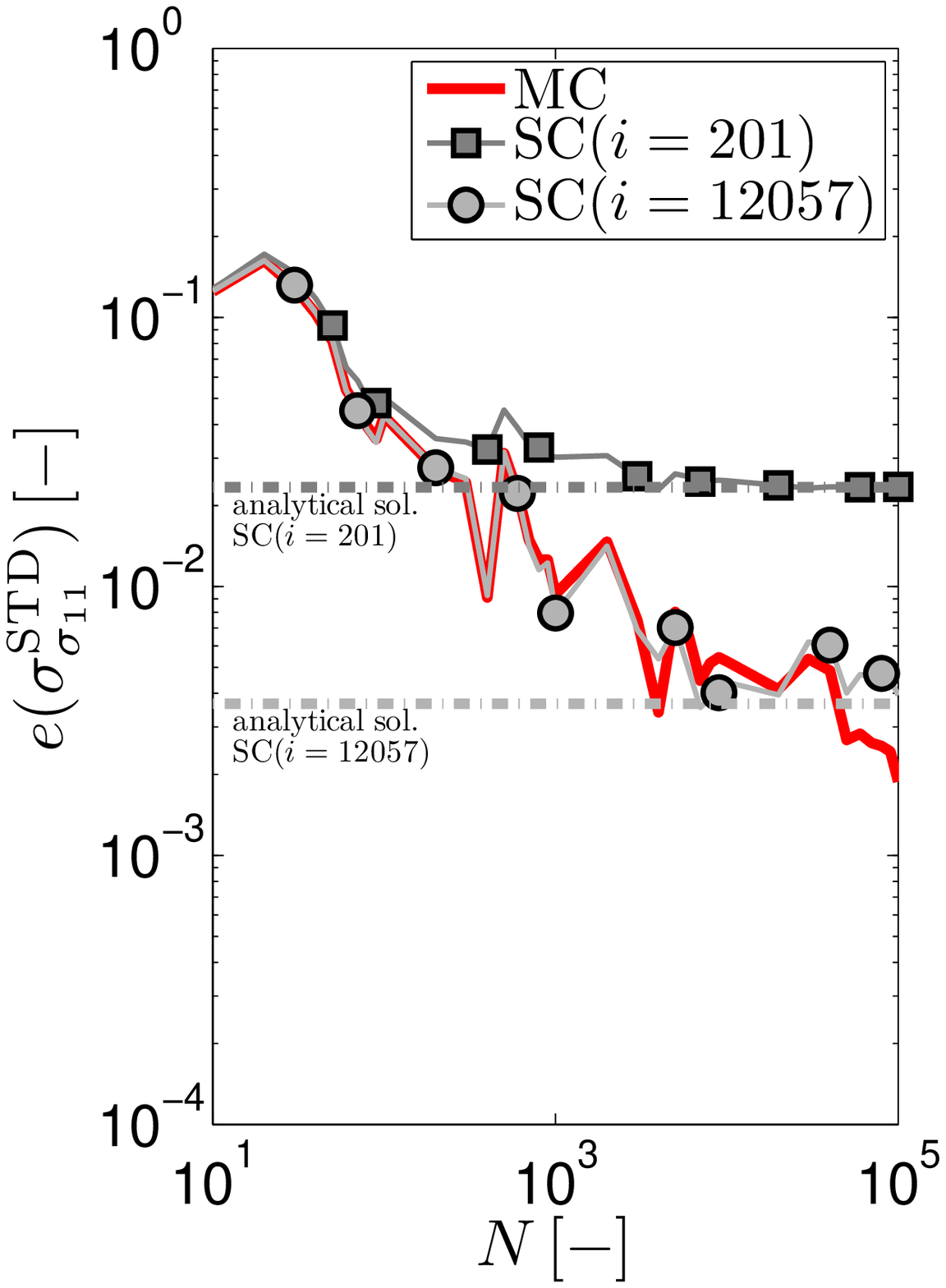} &
\includegraphics[keepaspectratio,width=4.5cm]{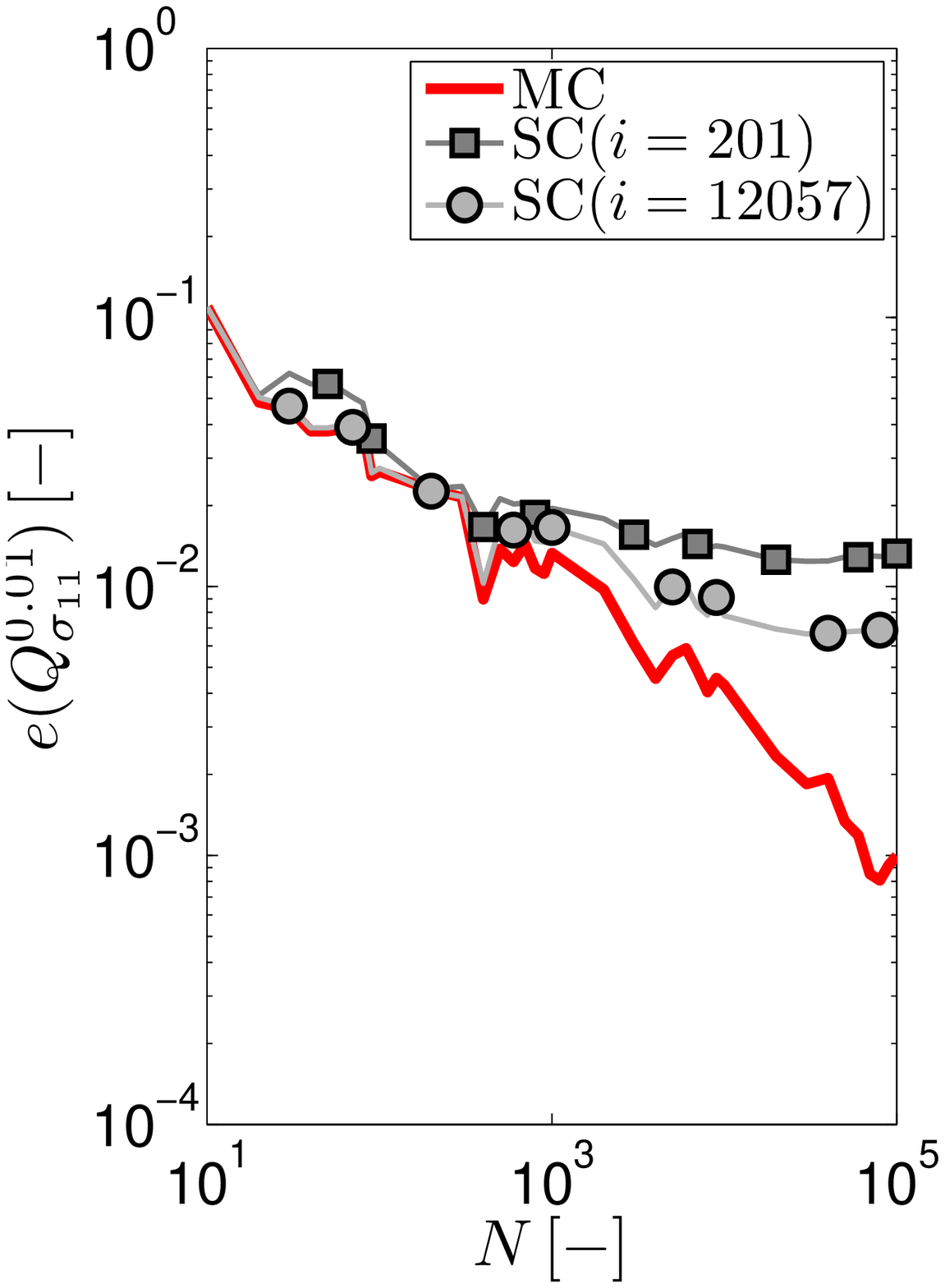}\\
(d) & (e) & (f) \\
\end{tabular}
%\caption{Error analysis of mean $\mu\,\mathrm{[-]}$ and standard
%deviation $\sigma^{\mathrm{STD}}\,\mathrm{[-]}$: (a) Fixed number
%of sparse grid points $(i=48231)$; (b) Fixed degree of polynomial
%chaos expansion $(p=5)$} \label{fig:err2}
\caption{Error analysis of (a) mean
$\mu\,\mathrm{[-]}$, (b)
  standard deviation $\sigma^{\mathrm{STD}}\,\mathrm{[-]}$ and (c) quantile $Q^{0.01}\,\mathrm{[-]}$ for fixed
  number of sparse grid points $(i=12057)$; (d) Error analysis of mean
  $\mu\,\mathrm{[-]}$, (e) standard deviation
  $\sigma^{\mathrm{STD}}\,\mathrm{[-]}$ and (f) quantile $Q^{0.01}\,\mathrm{[-]}$ for fixed degree of PCE $(p=5)$} \label{fig:err2}
\end{figure}

The Eqs. (\ref{eq:L21}) -- (\ref{eq:R2}) are utilised again to
asses the accuracy of surrogate model. Figs.~\ref{fig:err2}(a)-(b)
and Tab.~\ref{tab:err2} display the results of error analysis for
predicting mean $\mu$, standard deviation $\sigma^\mathrm{STD}$
and quantile $Q^{0.01}$ of a model response $\sigma_{11}$.  The
errors exhibit very similar convergence with increasing number of
collocation points and degree of polynomials as in previous
example.

\begin{figure}[h!]
\centering
\begin{tabular}{cc}
\includegraphics[keepaspectratio,width=6.5cm]{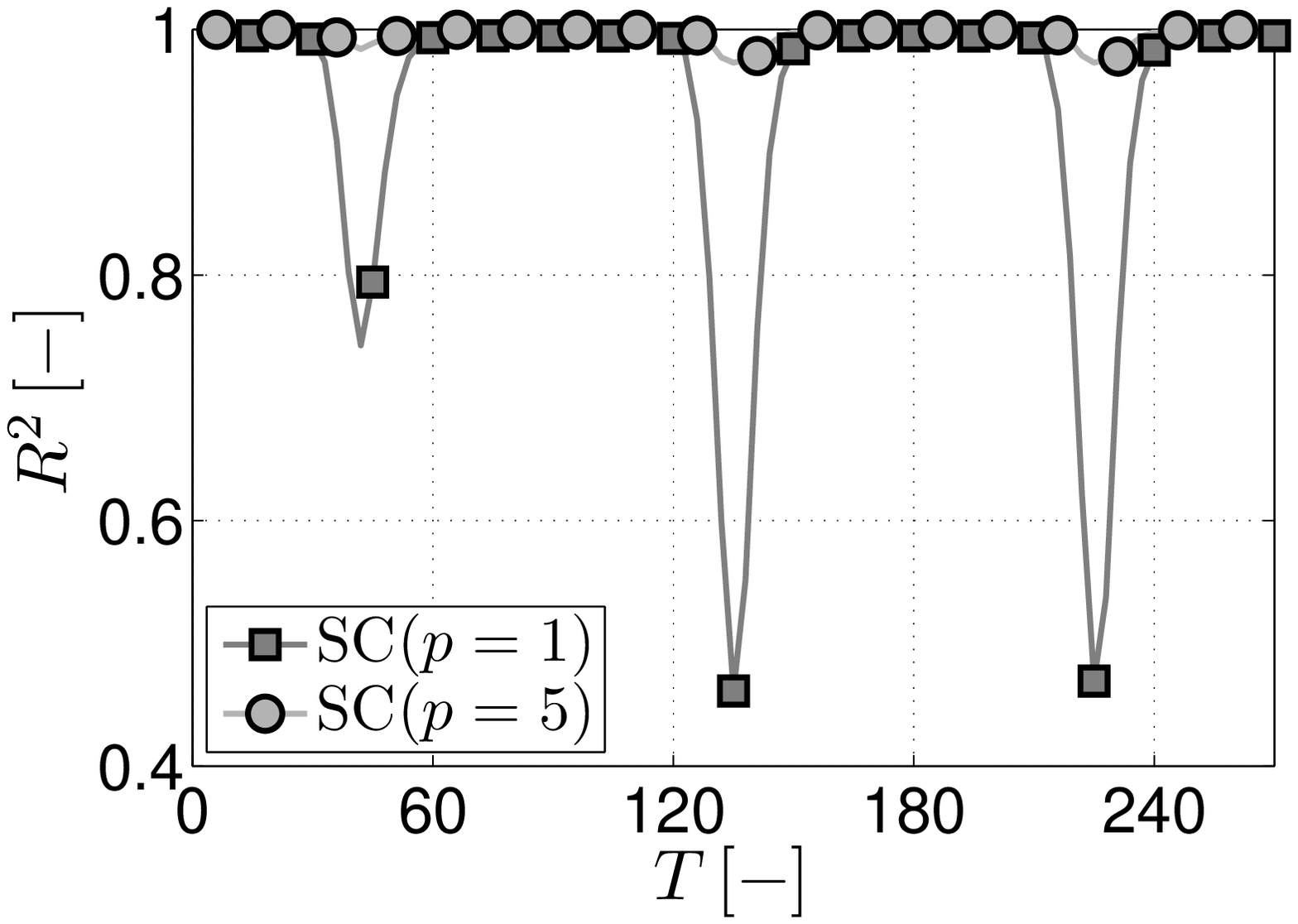} &
\includegraphics[keepaspectratio,width=6.5cm]{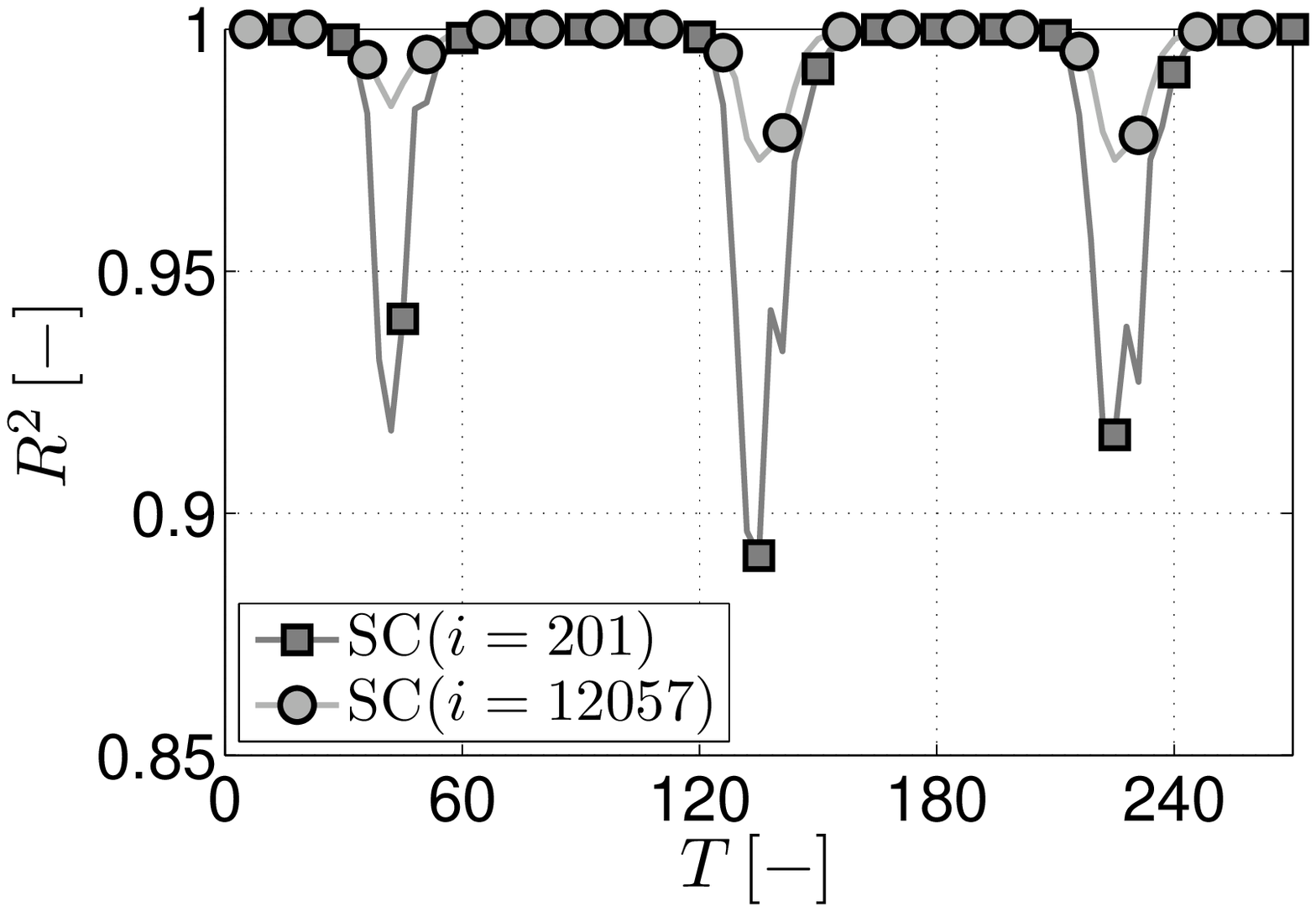}\\
(a) & (b) \\
\end{tabular}
\caption{Evolution of $R^2$: (a) Function of degree of PCE
($i=12057$); (b) Function of sparse grid points ($p=5$)}
\label{fig:R22}
\end{figure}

The coefficient of determination $R^2$ is plotted in
Figs.~\ref{fig:R22}(a)-(b) to quantify the overall accuracy of
surrogate models in particular time steps. The lower quality of PC
expansion is caused again by nonlinearity in the stochastic
solution near elastic limit of material.

\begin{figure}[t!]
\centering
\begin{tabular}{ccc}
\includegraphics[keepaspectratio,width=4.5cm]{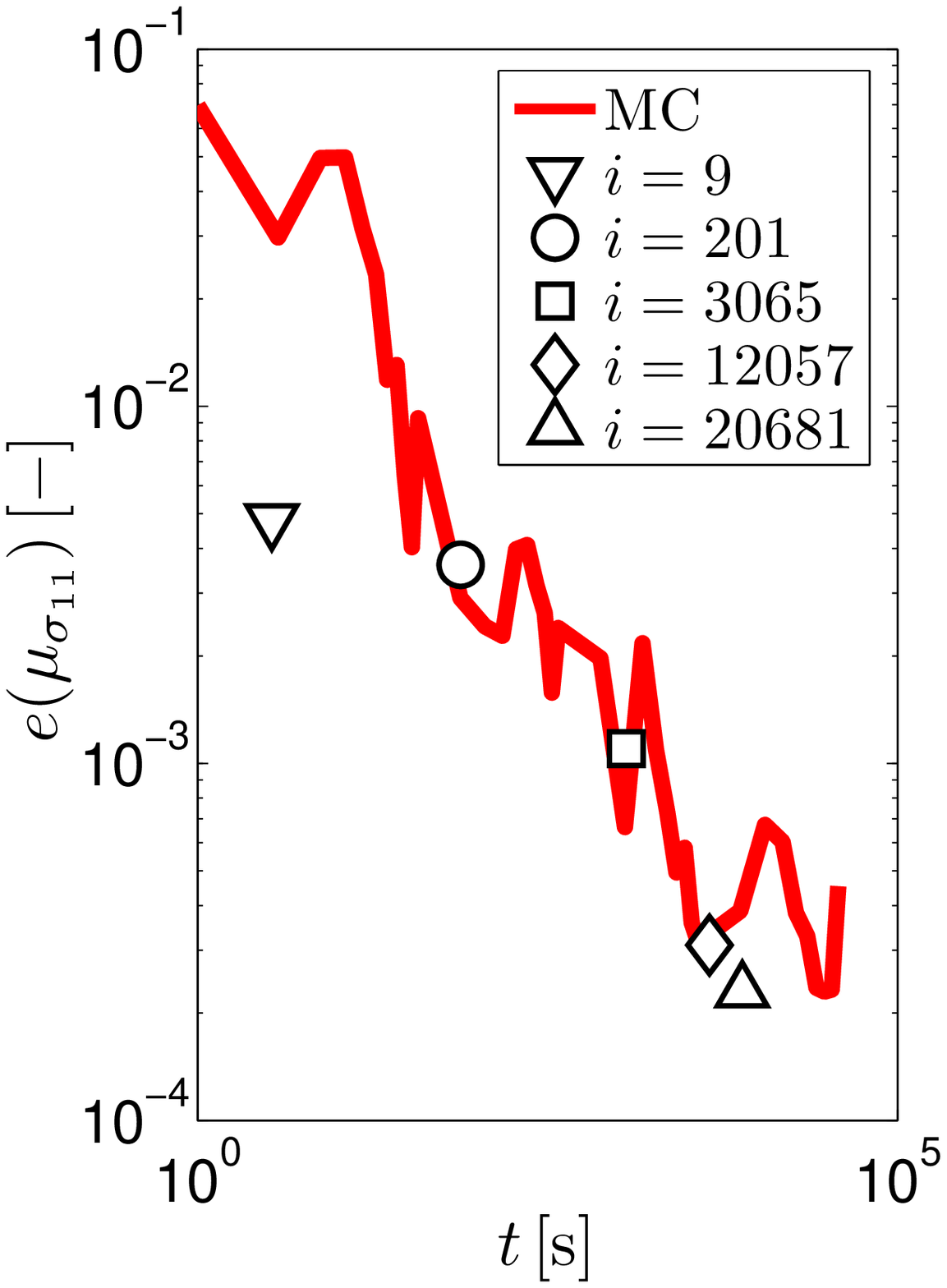} &
\includegraphics[keepaspectratio,width=4.5cm]{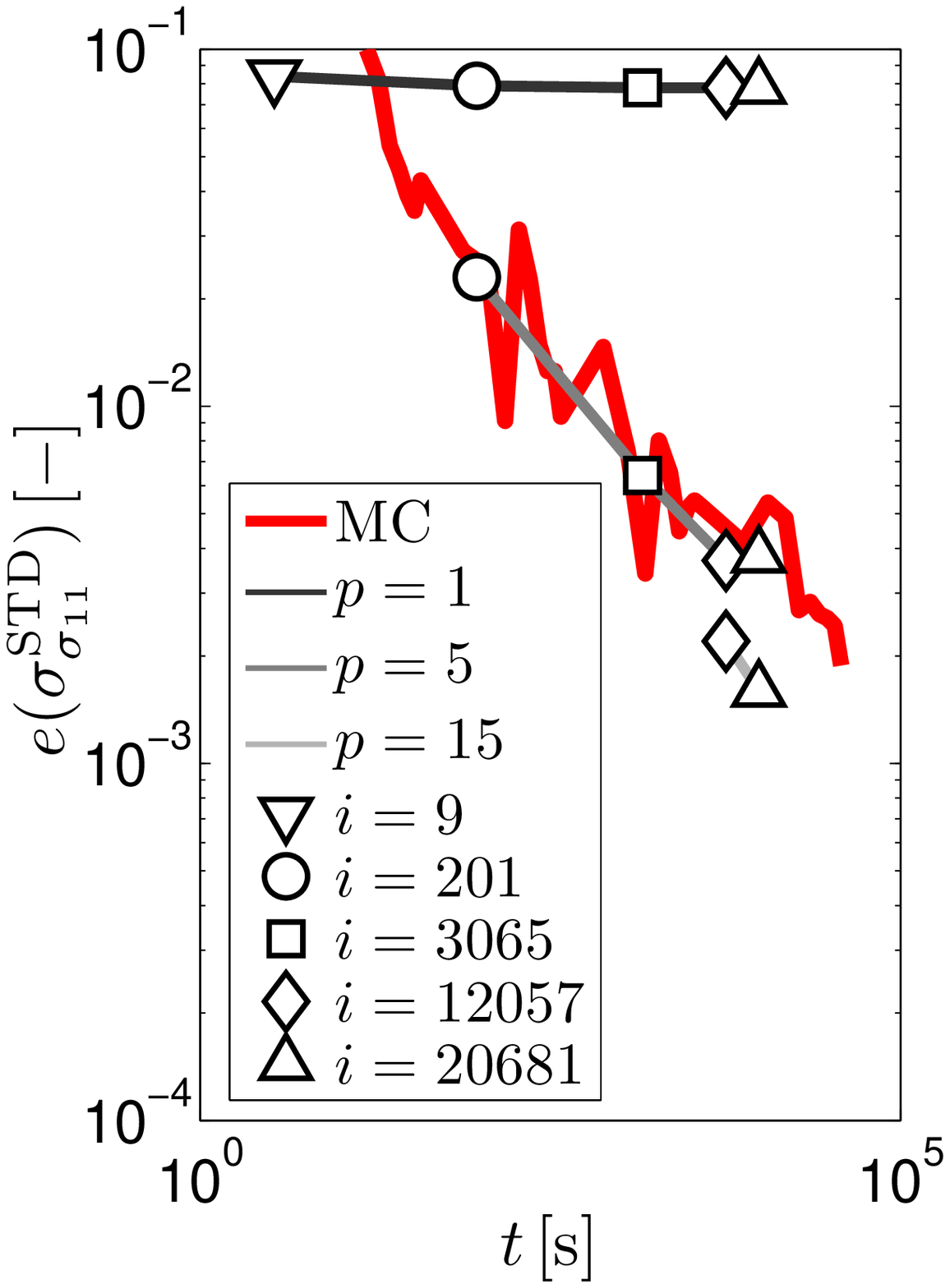} &
\includegraphics[keepaspectratio,width=4.5cm]{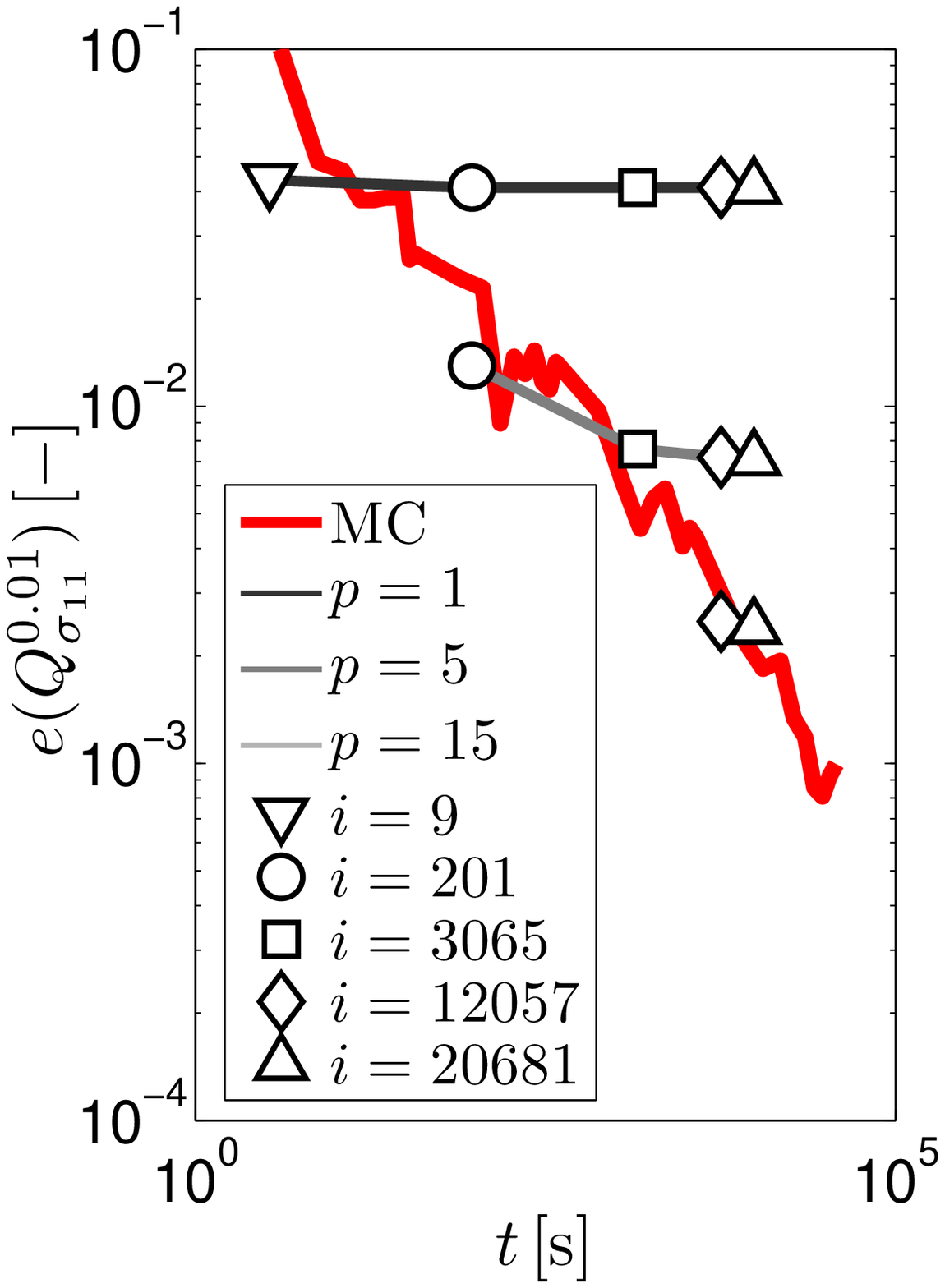} \\
(a) & (b) & (c) \\
\end{tabular}

\caption{Error analysis of (a) mean $\mu\,\mathrm{[-]}$, (b)
  standard deviation $\sigma^{\mathrm{STD}}\,\mathrm{[-]}$ and (c) quantile $Q^{0.01}\,\mathrm{[-]}$ as a function of computational time}
\label{fig:time}
\end{figure}

The evolution of time requirements as a function of number of
samples $N$ is presented in Fig.~\ref{fig:time} for MC
computations and PC-based computations performed for different
number of collocation points ($i=201$, $i=12057$). PC-based
predictions of mean $\mu$ and standard deviations
$\sigma^\mathrm{STD}$ are obtained analytically, while predictions
of quantile $Q^{0.01}$ are computed by MC sampling with $N = 10^5$
samples.

Computational time required for prediction of mean is composed
mainly by the time necessary for evaluation of collocation points,
because only the constant term of PCE needs to be calculated and
the corresponding computational time is negligible.
Fig.~\ref{fig:time}(a) shows that time requirements of MC-based
and PC-based predictions are comparable, which indicates that the
organisation of collocation points does not bring any significant
advantage comparing to randomicity of MC method. This may be
caused by high nonlinearity of the model, which needs to be
described by higher order polynomials constructed on higher number
of collocation points as suggested by better result of PC-based
predictions for $i=20681$ points.

Similar phenomenon may be observed also in Fig.~\ref{fig:time}(b),
where PC-based predictions outperform results of MC method also
for higher number of collocation points. Nevertheless, a
significant role is also played by the polynomial order used and
thus the computational time includes here also the time needed for
calculation of all the polynomial terms. This time is however
still negligible compared to the time needed for the evaluation of
collocation points as visible in Fig.~\ref{fig:time}(b). Here, the
points corresponding to the same number of collocation points but
different polynomial order have almost identical time coordinate.
This observation suggests an important conclusion and
recommendation to use the highest polynomial order possible for a
given number of collocation points -- in
Figs.~\ref{fig:time}(b)-(c), this situation correspond to left
ends of lines corresponding to particular polynomial orders.

The same conclusion may be done regarding Fig.~\ref{fig:time}(c),
even though here the computational time needed for PC-based
predictions includes also inevitable time for MC-sampling of the
polynomials with $N=10^5$ samples, which seems again negligible,
relatively compared to the time for evaluation of collocation
points. The number of MC samples of polynomials is chosen with
respect to Figs.~\ref{fig:err2}(c) and~\ref{fig:err2}(f), where
the curves of prediction errors seems to be stabilised between
$N=10^4$ and $N=10^5$ samples and higher number of samples will
probably not change the results. It is also interesting to point
out the negligible effect of increasing number of collocation
points for an unchanged polynomial order leading to waste of
computational time and worse results than MC-based predictions.

\section{Conclusions}
\label{sec:con}

This paper presents the numerical modelling of elasto-plastic
material behavior described by the uncertain parameters. In
particular, we employed the {\it Maxwell--Huber--Hencky--von
Mises} model, which is sufficiently robust to describe real-world
materials such as metals, but which is also nonlinear and
time-dependent material model. In order to investigate the
presence of uncertainty on material response, we replace the
expensive MC simulations of computational model by its cheaper
approximation based on PCE computed by SC method. The whole
concept is demonstrated on two simple examples of uniaxial test at
a material point, where interesting phenomena can be clearly
understood. First example consider only loading path and Young's
modulus to be uncertain, while the second example is extended by
load/unload cycle and uncertainty assumed in all the four material
parameters: Young's modulus, Poisson's ratio, initial yield
strength and hardening parameter. The quality of obtained
surrogate models is compared to MC method in terms of accuracy as
well as the time requirements. Figs.~\ref{fig:time}(a)-(c) show
that the second example represent situation where time
requirements are mainly driven by time needed for evaluation of
collocation points and PC construction is negligible. However, the
PC-based predictions achieve also similar accuracy as MC-based
predictions for equal number of sampling and collocation points as
demonstrated in Figs.~\ref{fig:err2}(a)-(f). Therefore, the
PC-based approximation does not bring an important acceleration in
this example. In the first example, the computational time of
model simulation is too small for any reliable measurement.
Nevertheless, we may point out the results in
Figs.~\ref{fig:err1}(a)-(f), where only $35$ collocation points
need to be computed for prediction of mean and standard deviation
with errors comparable to MC-based predictions obtained for more
than $10^3$ samples. Hence, the significant acceleration by
PC-based approximation cannot be guaranteed generally, but
definitely remains interesting and worthy to apply.

\section*{Acknowledgment}
This outcome has been achieved with the financial support of the
Czech Science Foundation, project No. 105/11/0411 and 105/12/1146.


\begin{thebibliography}{99}

\bibitem{Anders:1999}
Anders, ~M., Hori, ~M.: Stochastic finite element method for
elasto-plastic body, {\it International Journal for Numerical
Methods in Engineering}, {\bf 46}, 11, 1999, pp.~1897--1916.

\bibitem{Arnst:2012}
Arnst, ~M., Ghanem, ~R.: A variational-inequality approach to
stochastic boundary value problems with inequality constraints and
its application to contact and elastoplasticity, {\it
International Journal for Numerical Methods in Engineering}, {\bf
89}, 2012, pp.~1665--1690.

\bibitem{Babuska:2004}
Babu\v{s}ka, ~I., Tempone, ~R., Zouraris, ~G. E.: Galerkin Finite
Element Approximations of Stochastic Elliptic Partial Differential
Equations, {\it SIAM Journal on Numerical Analysis}, {\bf 42}, 2,
2004, pp.~800--825.

\bibitem{Babuska:2007}
Babu\v{s}ka, ~I., Nobile, ~F., Tempone, R.: A Stochastic
Collocation Method for Elliptic Partial Differential Equations
with Random Input Data. {\it SIAM Journal on Numerical Analysis},
{\bf 45}, 3, 2007, pp.~1005--1034.

\bibitem{Blatman:2010}
Blatman, ~G., Sudret, ~B.: An adaptive algorithm to build up
sparse polynomial chaos expansions for stochastic finite element
analysis, {\it Probabilistic Engineering Mechanics}, {\bf 25}, 2,
2010, pp.~183--197.

\bibitem{Dunne:2005}
Dunne, ~F., Petrinic, ~N.: {\it Introduction to Computational
Plasticity}, Oxford University Press, 2005.

\bibitem{Evans:1995}
Evans, ~M., Swartz, ~T.: Methods for approximating integrals in
statistics with special emphasis on Bayesian integration problems,
{\it Statistical Science}, {\bf 10}, 3, 1995, pp.~254--272.

\bibitem{Ghanem:2012}
Ghanem, ~R. G., Spanos, ~P. D.: {\it Stochastic Finite Elements: A
Spectral Approach}, Dover Publications, Revised edition, 2012.

\bibitem{Grassl:2006}
Grassl, ~P., Jir\'{a}sek, ~J.: Damage-plastic model for concrete
failure, {\it International Journal of Solids and Structures},
{\bf 43}, 2006, pp.~7166--7196.

\bibitem{Gutierrez:2004}
Guti\'{e}rrez, ~M., Krenk, ~S.: {\it Encyclopedia of Computational
Mechanics, chap. Stochastic finite element methods}, John Wiley \&
Sons, Ltd., 2004.

\bibitem{Heiss:2008}
Heiss, ~F., Winschel, ~V.: Likelihood approximation by numerical
integration on sparse grids, {\it Journal of Econometrics}, {\bf
144}, 1, 2008, pp.~62--80.

\bibitem{Horak:2009}
Hor\'{a}k, ~M.: {\it Localization Analysis of Damage and
Plasticity Models}, Master thesis, CTU in Prague, 2009.

\bibitem{Keese:2003}
Keese, ~A.: {\it A review of recent developments in the numerical
solution of stochastic partial differential equations (stochastic
finite elements)}, Tech. rep., Institute of Scientific Computing,
Technical University Braunschweig, 2003.

\bibitem{Krieg:1977}
Krieg, ~R.~D., Krieg, ~D.~B.: Accuracies of Numerical Solution
Methods for the Elastic-Perfectly Plastic Model, {\it Journal of
Pressure Vessel Technology}, {\bf 99}, 4, 1977, pp.~510--515.

\bibitem{Matthies:2005}
Keese, ~A., Matthies, ~H.~G.: Hierarchical parallelisation for the
solution of stochastic finite element equations, {\it Computers
and Structures}, {\bf 83}, 2005, pp.~1033--1047.

\bibitem{Kucerova:2012}
Ku\v{c}erov\'{a}, A., S\'{y}kora, J., Rosi\'{c}, B., Matthies,
~H.~G.: Acceleration of uncertainty updating in the description of
transport processes in heterogeneous materials, {\it Journal of
Computational and Applied Mathematics}, {\bf 236}, 18, 2012,
pp.~4862--4872.

\bibitem{Kucerova:2013}
Ku\v{c}erov\'{a}, A., S\'{y}kora, J.: Uncertainty updating in the
description of coupled heat and moisture transport in
heterogeneous materials, {\it Applied Mathematics and
Computation}, {\bf 219}, 2013, pp.~7252--7261.

\bibitem{Lourenco:1997}
Louren\c{c}o, ~P. B., de Borst, ~R., Rots, ~J. G.: A plane stress
softening plasticity model for orthotropic materials, {\it
International Journal for Numerical Methods in Engineering}, {\bf
40}, 1997, pp.~4033--4057.

\bibitem{Marzouk:2007}
Marzouk, ~Y., Najm, ~H., Rahn, ~L.: Stochastic Spectral Methods
for Efficient Bayesian Solution of Inverse Problems, {\it Journal
of Computational Physics}, {\bf 224}, 2, 2007, pp.~560--586.

\bibitem{Matthies:2007}
Matthies, ~H. G.: {\it Encyclopedia of Computational Mechanics,
chap. Uncertainty Quantification with Stochastic Finite Elements}.
John Wiley \& Sons, Ltd., 2007.

\bibitem{Rosic:2008}
Rosi\'{c}, B., Matthies, ~H.~G.: Computational Approaches to
Inelastic Media with Uncertain Parameters, {\it Journal of the
Serbian Society for Computational Mechanics}, {\bf 2}, 1, 2008,
pp.~28--43.

\bibitem{Rosic:2012}
Rosi\'c, B.: {\it Variational Formulations and Functional
Approximation Algorithms in Stochastic Plasticity of Materials},
PhD thesis, Institute of Scientific Computing, Technical
University Braunschweig, 2012.

\bibitem{Steel:1960}
Steel, ~R.~G.~D, Torrie, J.~H.: {\it Principles and Procedures of
Statistics with Special Reference to the Biological Sciences},
 McGraw Hill, 1960.

\bibitem{Stefanou:2009}
Stefanou, ~G.: The stochastic finite element method: Past, present
and future, {\it Computer Methods in Applied Mechanics and
Engineering}, {\bf 198}, 9-12, 2009, pp.~1031--1051.

\bibitem{Wiener:1938}
Wiener, ~N.: The Homogeneous Chaos, {\it American Journal of
Mathematics}, {\bf 60}, 4, 1938, pp.~897--936.

\bibitem{Xiu:2002}
Xiu, ~D., Karniadakis, ~G. E.: The Wiener–Askey Polynomial Chaos
for Stochastic Differential Equations, {\it SIAM Journal on
Scientific Computing}, {\bf 24}, 2, 2002, pp.~619--644.

\bibitem{Xiu:2005}
Xiu, ~D., Hesthaven, ~J. S.: High-Order Collocation Methods for
Differential Equations with Random Inputs, {\it SIAM Journal on
Scientific Computing}, {\bf 27}, 3, 2005, pp.~1118--1139.

\bibitem{Xiu:2009}
Xiu, ~D.: Fast Numerical Methods for Stochastic Computations: A
Review, {\it Communications in Computational Physics}, {\bf 5},
2-4, 2009, pp.~242--272.

\end{thebibliography}
\end{document}